\documentclass[pra,aps,twocolumn,superscriptaddress,showpacs,10pt]{revtex4-1}
\usepackage{amsmath,amssymb,graphicx}

\begin{document}

\title{Dissipatons: Stationary and stable light beam propagation in Kerr media with nonlinear absorption with controllable dissipation patterns}

\author{Miguel A. Porras}
\affiliation{Grupo de Sistemas Complejos, Universidad Polit\'ecnica de Madrid, ETSIM, Rios Rosas 21, 28003 Madrid, Spain}
\author{Carlos Ruiz-Jim\'enez}
\affiliation{Grupo de Sistemas Complejos, Universidad Polit\'ecnica de Madrid, ETSIA, Ciudad Universitaria s/n, 28040 Madrid, Spain}
\author{M\'arcio Carvalho}
\affiliation{Grupo de Sistemas Complejos, Universidad Polit\'ecnica de Madrid, ETSIA, Ciudad Universitaria s/n, 28040 Madrid, Spain}

\begin{abstract}
We report on the stationary and robust propagation of light beams with rather arbitrary and controllable intensity and dissipation transverse patterns in self-focusing Kerr media with nonlinear absorption. When nonlinear absorption is due to multi-photon ionization at high beam powers in transparent media such as glasses or air, these beams can generate multiple plasma channels with tailored geometries. Their nature and spatial characteristics are discussed in detail, as well as the laws determining their spontaneous formation from coherent superpositions of Bessel beams of different amplitudes and topological charges.
\end{abstract}


\maketitle

\section{Introduction}

The achievement of localized, stationary and stable propagation of multidimensional waves in homogeneous, isotropic nonlinear media is one of the main problems in nonlinear optics, and constitutes today a broad area of research \cite{MALOMED1,KIVSHAR,CHEN}. Many of the elaborated solutions, some of them realized experimentally, rely on two broad models. The first one is the conservative model based on nonlinear Schr\"odinger equations with dispersive, Kerr-type nonlinearities and others, describing passive propagation of intense waves in unbounded transparent media. The second model is the complex Ginzburg-Landau equation in systems with gain and loss, as laser cavities \cite{AKHMEDIEV,ACKEMANN,MALOMED2}. More or less broad families of solitons and dissipative solitons (also called auto-solitons or cavity-solitons) with a variety of geometries have been reported to be supported by these models. Given their relation with the present article, we mention the families of vortex-solitons ---solitons with a nested vortex--- \cite{KRUGLOV}, their azimuthally modulated versions called azimuthons \cite{DESYATNIKOVPrl1,MINOVICHOe} and solitons clusters \cite{DESYATNIKOVPrl2}, see \cite{DESYATNIKOVReview} for a review. In the conservative models, a stable balance between diffraction and the dispersive nonlinearities supports the robust stationary propagation. In the dissipative model, an additional balance is needed between gain and losses \cite{PAULAU,CRASOVAN,ZHANGPra}.

Non-conservative models involving only losses are usually left aside because they do not support solitons, in the sense of self-trapped structures. However the nonlinear Schr\"odinger equation with purely loss terms and no gain terms arises naturally in situations as common as light beam propagation in transparent dielectrics media at intensities such that multi-photon absorption processes are relevant. The existence of light beams, called nonlinear unbalanced Bessel beams, with the ability to propagate without any change, including no attenuation, in media with Kerr-type nonlinearities and nonlinear absorption, was described in 2004 \cite{PORRASPrl}. They provided a theoretical basis for interpreting some features of light filaments generated from intense, ultrashort pulses \cite{COUAIRONPhysRep,BERGERep,BERGEDis}, where nonlinear absorption arises from multi-photon ionization, particularly the replenishment and self-healing mechanisms involved in their robust and quasi-stationary propagation  \cite{DUBIETIS,GAIZAUSKAS,FACCIOOptComm,ROSOAppPhysA,ZEMLYANOV,DUBIETISLithuanian,FACCIOPra}. Stationarity with nonlinear absorption is indeed possible for less strongly localized waves than solitons, as nonlinear versions of Bessel-like and Airy-like beams \cite{DURNINPrl,SIVILOGLOUOptLett} carrying, ideally, an infinite amount of power, or a power reservoir. In these beams \cite{PORRASPrl,PORRASJosaB,JUKNAOe,XIESci,LOTTIPra} the power in the reservoir is permanently flowing towards the central region of high intensity, where most of nonlinear power losses take place, replenishing it. After the fundamental nonlinear unbalanced Bessel beam \cite{PORRASPrl}, the same mechanism has been more recently seen to work with high-order (vortex-carrying) nonlinear Bessel beams \cite{PORRASJosaB,JUKNAOe,XIESci}. Here we will refer to all them as nonlinear Bessel vortex beams (nonlinear BVBs), the fundamental beam being the particular vortex-less situation.

Nonlinear BVBs have proven to be particularly fruitful to explain the filamentation dynamics seeded by ultrashort (linear) Bessel beams \cite{POLESANAPra,AKTURKOptComm,POLESANAPre,POLESANAPrl,POLESANAOe,MAJUSJosaB,AKTURKSpie,KAYAAip,JUKNAOe,XIESci}. Even if the physics of filamentation involves complex light-matter interactions and rich spatiotemporal dynamics \cite{COUAIRONPhysRep,BERGERep,BERGEDis}, the three key phenomena determining substantially the filamentation dynamics with conical beams are just those supporting the stationarity of nonlinear BVBs, namely, diffraction, Kerr self-focusing and the nonlinear absorption \cite{POLESANAPra,JUKNAOe,XIESci}. The steady and unsteady regimes of filamentation observed in the filamentation with the lowest-order Bessel beam \cite{POLESANAPra} have been successfully explained in terms of nonlinear BVB attractors and their stability properties \cite{PORRASPra1}. More recently, tubular filamentation from vortex-carrying Bessel beams has been created \cite{JUKNAOe,XIESci}, and has also been interpreted as the formation of a vortex-carrying nonlinear BVB that is stable against radial and azimuthal perturbations \cite{PORRASPra2}. The disintegration of the tubular regime into rotatory or random filaments observed in related experiments is also explained in terms of the development of the azimuthal instability of unstable nonlinear BVBs \cite{PORRASPra2}.

The intensity profiles of nonlinear BVBs are all circularly symmetric, and so they are the dissipation channels that they can generate. The tubular regime is a sophisticated example of a ``tailored" or controlled transversal pattern, which is expected to open new perspectives in laser-powered material processing and other applications \cite{XIESci,COURVOISIERApplPhys,BHUYAN,KRISNALaserPhysics,ARNOLDJPhysB,COURVOISIER2}. Complex filamentation patterns other than circular arise naturally in multiple filamentation from small-scale self-focusing of laser beams carrying many critical powers \cite{COUAIRONPhysRep,BERGERep,BERGEDis}, or from the azimuthal instability of tubular filaments \cite{JUKNAOe,XIESci,PORRASPra2}, but originating from noisy perturbations, these patterns are largely uncontrollable and far from being propagation-invariant.

In this paper we investigate if propagation-invariant light beams with other geometries, preferably with controllable intensity and dissipation patterns, are supported by transparent dielectrics with self-focusing nonlinearities and nonlinear absorption. We find the affirmative answer when studying by means of the numerical simulations the nonlinear propagation dynamics seeded by illuminating the medium with coherent superpositions of Bessel beams of different topological charges and amplitudes. These (linear) Bessel beam superpositions are easily generated using spatial light modulators \cite{FROEHLYJosaA,CLARKOe,VALILYEUOe}, illuminating an axicon with several coherent Gaussian beams with embedded vorticities, or with superpositions of Laguerre-Gauss beams \cite{ARLTOptCommun}. For a range of the parameters that define the input Bessel beam superposition and the material medium,  stationary and stable propagation states dissipating constantly their power along organized and narrow channels are seen to emerge spontaneously. We confine ourselves to the superposition of two Bessel beams, where the channels geometry is primarily controlled by their amplitudes and topological charges, and the variety of intensity and dissipation patterns is already very vast.

To clearly distinguish these states from other known localized structures, we first present the more general states without any particular symmetry, and then we focus on those with $n$-fold rotational symmetry formed from superpositions of two Bessel beams of opposite topological charges, resembling azimuthons \cite{DESYATNIKOVPrl1,MINOVICHOe,PAULAU,CRASOVAN,ZHANGPra}. These structures are nevertheless neither conservative nor supported by a gain-loss balance, and do not rotate either. Given the distinctive properties of having arbitrary shapes, and of not being supported by gain, i. e., of being strictly dissipative, we have called them ``dissipatons".

Outside their central region, dissipatons are always seen to have linear conical tails, and therefore a power reservoir, shaped like a superposition of unbalanced Bessel beams (i. e., with unequal inward and outward H\"ankel beam amplitudes), all them with the same cone angle but different topological charges. The excess of the inward H\"ankel beam amplitudes creates an inward radial current, as that in nonlinear BVBs but with an azimuthal dependence, and such that net power flux coming from the reservoir equals the total power loss rate in the dissipative core. We find simple rules that describe the dissipaton formation. Given an input superposition of Bessel beams, the dissipaton that emerges preserves the cone angle, the topological charges and the amplitudes of the inward H\"ankel components of all input Bessel beams. Although the dissipative center has a more complex structure of intertwined hot spots and vortices, it reflects up to certain extent the structure of the linear tails, and therefore that of the input Bessel beam superposition, the dissipation pattern being then quite controllable. Numerical simulations also show that dissipatons are stable under large perturbations strongly distorting the dissipative center. In fact, the only condition for the whole dissipaton structure to emerge spontaneously is the presence of the inward currents carried by its inward H\"ankel beam components.

\section{Basic relations}\label{BASIC}

According to the above, we consider the propagation of a paraxial light beam $E=\mathrm{Re}\left\{ A\exp [i(kz-\omega t)]\right\} $, with carrier frequency $\omega$,  propagation constant $k=n\omega /c$ ($n$ is the linear refractive index and $c$ the speed of light in vacuum) and with complex envelope $A$. We assume that cubic Kerr nonlinearity and nonlinear absorption are the dominating nonlinearities of the medium. As with nonlinear BVBs, other dispersive nonlinearities (quintic, saturable) could be considered as well without substantial changes in the results. The evolution of the envelope is then ruled by the nonlinear Schr\"odinger equation (NLSE)
\begin{equation}\label{NLSE}
\partial _{z}A=\frac{i}{2k}\Delta_\perp A+i\frac{kn_{2}}{n}|A|^{2}A-\frac{\beta ^{(M)}}{2}|A|^{2M-2}A\,,
\end{equation}
where $\Delta _\perp$ is the transverse Laplacian given by $\Delta_\perp =\partial _{r}^{2}+(1/r)\partial _{r}+(1/r^{2})\partial _{\varphi }^{2}$ in cylindrical coordinates $(r,\varphi,z)$, $n_2>0$ is the nonlinear refractive index, and $\beta ^{(M)}>0$ is the $M$-photon-absorption coefficient.

At low enough intensities (neglecting the nonlinear terms), Eq. (\ref{NLSE}) is satisfied by the linear (and paraxial) Bessel beams $A(r,\varphi,z)\propto J_s(k\theta r)e^{is\varphi}e^{i\delta z}$ of cone angle $\theta$, with shortened axial wave vector by $\delta =-k\theta^2/2 <0$, and carrying a vortex of topological charge $s$. To investigate on nonlinear beams characterized also by a cone angle $\theta$ but at high intensities, we find it convenient to use the scaled and dimensionless radial coordinate, propagation distance and envelope
\begin{equation}  \label{SCALING}
\rho \equiv k\theta r=\sqrt{2k|\delta |}r, \quad \zeta \equiv |\delta |z,\quad \tilde{A}\equiv \left(\frac{\beta^{(M)}}{2|\delta |}\right)^{\frac{1}{2M-2}}A
\end{equation}
[in many Figures below, we will also use the dimensionless cartesian coordinates $(\xi,\eta) \equiv k\theta (x,y)$], to rewrite Eq. (\ref{NLSE}) as
\begin{equation}
\partial_\zeta\tilde{A}=i\Delta_\perp \tilde{A}+i\alpha |\tilde{A}|^{2}\tilde{A}-|\tilde{A}|^{2M-2}\tilde{A}\,,  \label{NLSE2}
\end{equation}
where now $\Delta_\perp = \partial _\rho^2+(1/\rho)\partial _{\rho}+(1/\rho^2)\partial _{\varphi}^2$, and
\begin{equation}  \label{alpha}
\alpha \equiv \left( \frac{2|\delta |}{\beta ^{(M)}}\right) ^{1/(M-1)}\frac{k n_{2}}{n|\delta |}\, .
\end{equation}
With these variables, the linear Bessel beams reads as $\tilde A \propto J_s(\rho)e^{is\varphi}e^{-i\zeta}$. Values of $|\tilde A|^2$ of the order of unity correspond to the typical values of the intensities involved in filamentation experiments in solids or gases, e. g., at $800$ nm with cone angle $\theta=1^\circ$ in water ($n=1.33$, $n_2=2.6\times 10^{-16}$ cm$^2$/W, $M=5$, and $\beta^{(5)}=8.3\times 10^{-50}$ cm$^7$/W$^4$ \cite{COUAIRONSpie}), or $0.1^\circ$ in air ($n\simeq 1$, $n_2=3.2\times 10^{-19}$ cm$^2$/W, $M=8$, and $\beta^{(8)}=1.8\times 10^{-94}$ cm$^{13}$/W$^7$ \cite{COUAIRONSpie}), $|\tilde A|^2\simeq 1$ corresponds to $4$ TW/cm$^2$ and $20$ TW/cm$^2$, respectively. The value of the parameter $\alpha$ is determined by the cone angle and the medium properties at the light wavelength. At $800$ nm in water, the values of $\alpha$ range from $0.17$ to $5.3$ for cone angles $\theta$ from $10^\circ$ to $1^\circ$, or in air, $\alpha$ varies between $0.08$ and $4.2$ for $\theta$ varying between $1^\circ$ and $0.1^\circ$.

We will further write the complex envelope as $\tilde A=\tilde a e^{i\Phi}$, where $\tilde a>0$ and $\Phi$ are its real amplitude and phase. The NLSE (\ref{NLSE2}) is then seen to be equivalent to the energy transport and eikonal equations
\begin{align}
&\frac{1}{2}\partial_\zeta \tilde a^2+ \nabla_\perp \cdot \mathbf j =-\tilde a^{2M}\, , \label{A} \\
&\partial_\zeta \Phi = \frac{\Delta_\perp \tilde a}{\tilde a} - \left|\nabla_\perp \Phi\right|^2 +\alpha \tilde a^2\, , \label{PHASE}
\end{align}
where $\nabla_\perp=\mathbf u_\rho \partial_\rho + \mathbf u_\varphi(1/\rho)\partial_\varphi$, $\mathbf u_\rho$ and $\mathbf u_\varphi$ are unit radial and azimuthal vectors, and $\mathbf j =\tilde a^2 \nabla_\perp \Phi=\mbox{Im}\{\tilde A^\star \nabla_\perp \tilde A\}$ is the current of the intensity $\tilde a^2$ in the transversal plane. Nonlinear BVBs were described as particular solutions to Eq. (\ref{A}) and Eq. (\ref{PHASE}) that generalize Bessel beams to nonlinear media, whose intensity profiles are propagation-invariant and circularly symmetric \cite{PORRASPrl,PORRASJosaB,JUKNAOe,XIESci}. Here we search for conical beams of the most general form $\tilde A=\tilde a(\rho,\varphi) e^{i\phi(\rho,\varphi)} e^{-i\zeta}$, whose intensity profiles $\tilde a^2(\rho,\varphi)$ are also localized but may no have any particular symmetry. According to Eqs. (\ref{A}) and (\ref{PHASE}), their amplitude and phase patterns, $a(\rho,\varphi)$ and $\phi(\rho,\varphi)$, must satisfy
\begin{align}
&\nabla_\perp \cdot \mathbf j = -\tilde a^{2M}\, , \label{DIV}\\
&\Delta_\perp \tilde a-\left|\nabla_\perp \phi\right|^2\tilde a + \tilde a + \alpha \tilde a^3 =0\, \label{AMP} ,
\end{align}
where $\mathbf j=\tilde a^2\nabla_\perp \phi$ is the $\zeta$-independent intensity current. Equation (\ref{DIV}) expresses the general condition for a conical beam to propagate without any change in a medium with absorption. Integrating over an arbitrary area $S$ in the transversal plane, and using the divergence theorem, this condition is readily seen to be equivalent to
\begin{equation}\label{C}
-\oint_{C} \mathbf j \cdot \mathbf n \; dl =\int_S \tilde a^{2M} ds\, ,
\end{equation}
or $-F_{C} =N_S$ for short, where  $dl$ is a differential element of the contour $C$ of $S$, $\mathbf n$ is the outward normal unit vector, and $ds$ and elemental area of $S$. This condition thus establishes that the power loss rate $N_S$ in any region $S$ of the transversal plane is offset by an inward power flux $-F_C$ though its contour for the stationary propagation to be possible. In particular, applying Eq. (\ref{C}) to a circle of large enough radius $\rho$ for the total nonlinear power losses to occur inside it (assumed they are finite), this condition requires
\begin{equation}\label{CR}
F_\infty \equiv -\int_0^{2\pi} j_\rho (\rho,\varphi) \rho d\varphi = N_\infty\, ,
\end{equation}
for $\rho\rightarrow \infty$, i. e., an inward power flux from the power reservoir of the conical beam that is independent of the large radius $\rho$, and that replenishes the power losses,
\begin{equation}
N_\infty \equiv \int_{\mathbb{R}^2}\tilde a^{2M}ds\, ,
\end{equation}
in the entire transversal section.

A direct search of solutions of the nonlinear partial differential equations (\ref{DIV}) and (\ref{AMP}) without simplifying assumptions, such as the circular symmetry, separability of the intensity and phase patterns in $\rho$ and $\varphi$, or particular boundary conditions, is complicated. An alternate procedure is suggested by a well-established connection between linear Bessel beams and nonlinear BVBs in the circularly symmetric case, as summarized below.

\section{Circularly symmetric case: Nonlinear Bessel vortex beams} \label{NBVB}

\begin{figure}[t!]
\begin{center}
\includegraphics*[width=4cm]{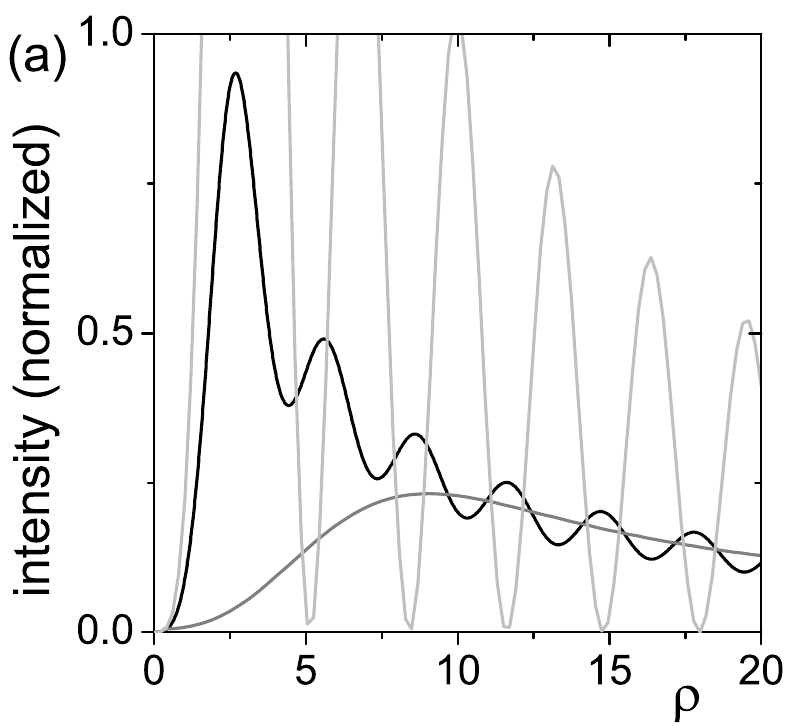}\includegraphics*[width=3.8cm]{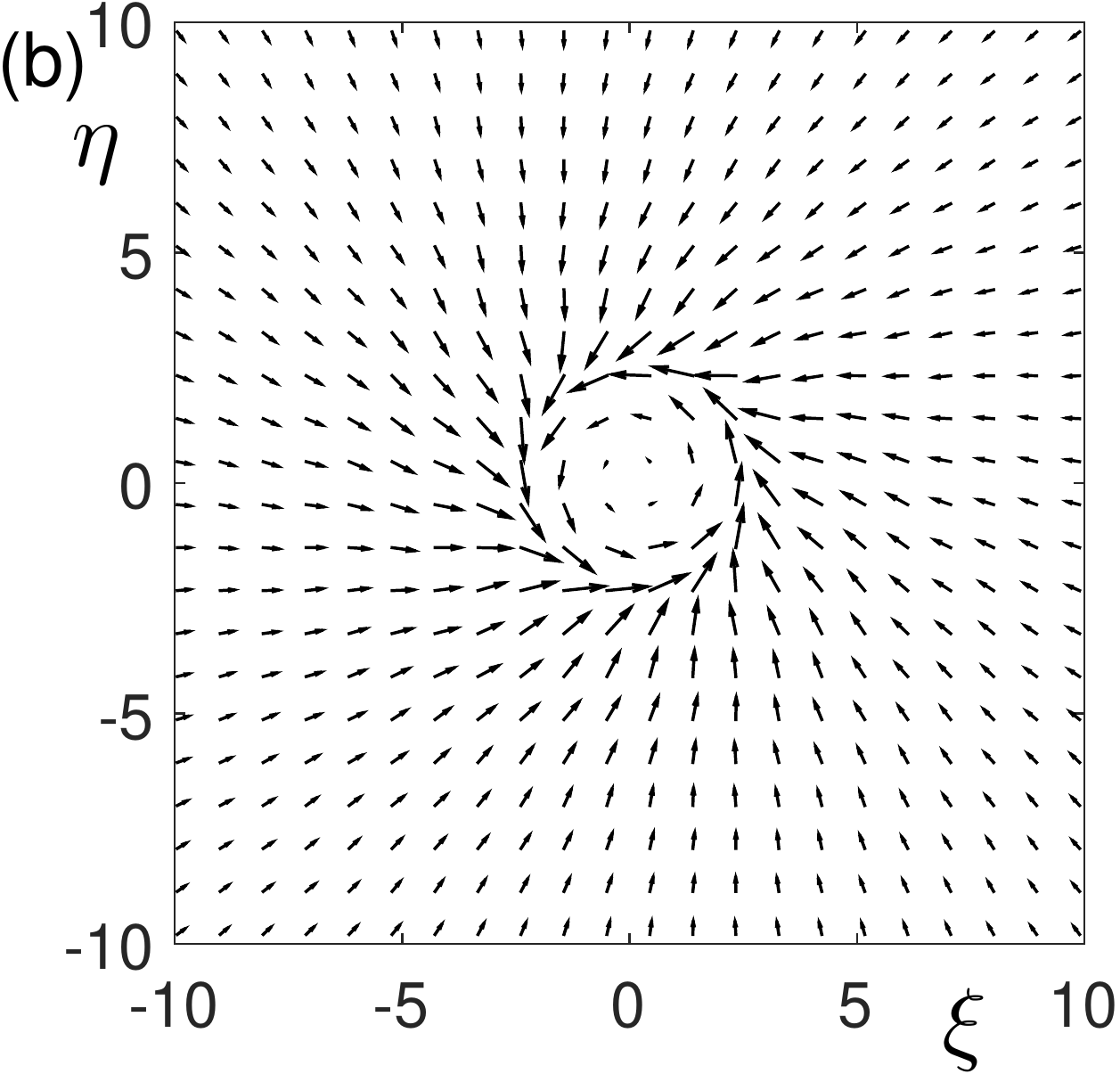}\\
\includegraphics[width=8cm]{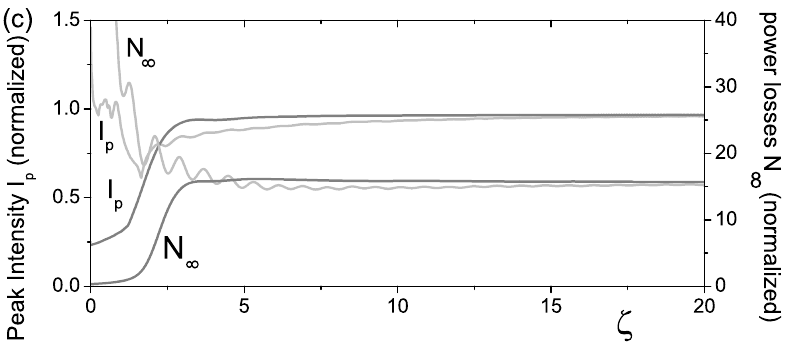}
\end{center}
\caption{\label{Fig1} (a) Intensity of the initial Bessel beam $\tilde A(\rho,\varphi,\zeta=0)= b_L J_s(\rho)e^{is\varphi}$, $b_L=4$, $s=2$ (light gray curve), of the initial ``holed" H\"ankel beam $\tilde A(\rho,\varphi,\zeta=0)= \frac{1}{2} b_L H^{(2)}_s(\rho)e^{is\varphi}(1-e^{-\rho^2/\rho_0^2})$, $b_L=4$, $s=2$, hole radius $\rho_0=6$, with the same asymptotic inward intensity current (dark gray curve), and the stationary intensity profile of the attracting nonlinear BVB formed at $\zeta=20$ (black curve) in a medium and with cone angle such that $M=5$ and $\alpha=0.5$. (b) Intensity current in the transversal plane of the attracting nonlinear BVB. (c) Peak intensity and nonlinear power losses versus propagation distance for the input Bessel beam (light gray curves) and for the holed H\"ankel beam (dark gray curves). All quantities are normalized as explained in the text. Normalized cartesian coordinates are $(\xi,\eta)=k\theta(x,y)$.}
\end{figure}

If a powerful Bessel beam is introduced into the medium, its nonlinear dynamics is determined by the existence and properties of an attracting nonlinear BVB \cite{PORRASPra1,PORRASPra2}. Given the input Bessel beam $\tilde A(\rho,\varphi,\zeta=0)=b_L J_s(\rho)e^{is\varphi}$, or equivalently $\tilde A(\rho,\varphi,\zeta=0)=\frac{1}{2}\left[b_L H_s^{(1)}(\rho)+ b_LH_s^{(2)}(\rho)\right]e^{is\varphi}$, the attracting nonlinear BVB, $\tilde a(\rho)e^{i\phi(\rho)}e^{is\varphi}e^{-i\zeta}$, has the same cone angle and topological charge as the linear Bessel beam, and behaves asymptotically at large radius $\rho$, where it remains linear, as the unbalanced Bessel beam $\tilde A\simeq \frac{1}{2}\left[b_{\rm out} H_s^{(1)}(\rho)+ b_{\rm in}H_s^{(2)}(\rho)\right]e^{is\varphi}e^{-i\zeta}$, with the property that $|b_{\rm in}|=b_L$ \cite{PORRASJosaB,PORRASPra1,PORRASPra2}. The conservation of the cone angle, the topological charge and the amplitude of the inward H\"ankel beam component specifies unambiguously an attracting nonlinear BVB for any input Bessel beam. Using the equivalent expressions
\begin{equation}\label{HANKEL}
H_s^{(1,2)}(\rho)\simeq\sqrt{\frac{2}{\pi\rho}}e^{\pm i(\rho-\pi/4-\pi s/2)}\, ,
\end{equation}
of the H\"ankel functions at large $\rho$, the outward H\"ankel beam, $(b_{\rm out}/2)H_s^{(1)}(\rho)e^{is\varphi}e^{-i\zeta}$ is seen to carry the current $\mathbf j =(|b_{\rm out}|^2/2\pi\rho)[\mathbf u_\rho + (s/\rho)\mathbf u_\varphi]$ spirally outwards, while the inward H\"ankel beam $b_{\rm in}H_s^{(2)}(\rho)e^{is\varphi}e^{-i\zeta}$, the current $\mathbf j =(|b_{\rm in}|^2/2\pi\rho)[- \mathbf u_\rho + (s/\rho)\mathbf u_\varphi]$ spirally inwards. From the asymptotic form of nonlinear BVBs and Eq. (\ref{CR}), one then gets the relation $-F_\infty =|b_{\rm in}|^2 - |b_{\rm out}|^2=N_\infty$ for the inward power flux restoring the total power losses in the nonlinear BVB.

The particular Bessel beam dynamics depends on the stability properties of the attracting nonlinear BVB. According to the linear-stability analysis in Refs. \cite{PORRASPra1,PORRASPra2}, there exist nonlinear BVBs that are stable against radial and azimuthal perturbations. In case of stability, the Bessel beam dynamics ends in the complete formation of the nonlinear BVB. The example of Fig. \ref{Fig1} summarizes these previous results. The input Bessel beam represented by the light gray curve in Fig. \ref{Fig1}(a) transforms into the nonlinear BVB represented by the black curve, having a radial inward component of the intensity current that sustains the stationarity, as seen in Fig. \ref{Fig1}(b). After a transient, the peak intensity and nonlinear power losses $N_\infty$ reach the constant values corresponding to the attracting nonlinear BVB [light gray curves in Fig. \ref{Fig1}(c)].

The fact that the linearly-stable nonlinear BVB is formed starting from so distant initial condition as the linear Bessel suggests stability under  large perturbations within a certain attraction basin. Additional numerical simulations indicate that the same final nonlinear BVB is formed when the input beam carries at large radius only the inward H\"ankel beam component that is preserved in the propagation. The attraction basin of the nonlinear BVB then includes all beams of the type $\tilde A = \frac{1}{2} b_L H_s^{(2)}(\rho)e^{is\varphi} + f(\rho,\varphi)$, where $f(\rho,\varphi)$ does not carry any additional inward radial current at large radius. The dark gray curves in Figs. \ref{Fig1}(a) and (c) show the radial intensity profile of an input beam with these characteristics that transforms into the same nonlinear BVB, and the peak intensity and power losses reaching the same constant values as with the input Bessel beam.

On the opposite side, if the attracting nonlinear BVB determined by the conservation of the cone angle, the topological charge and the amplitude of the inward H\"ankel beam component turns out to be unstable (according to the linear-stability analysis), the Bessel beam dynamics consists on the growth of its unstable modes, developing into large, periodic or chaotic oscillations around the nonlinear BVB if radial instability is dominant \cite{PORRASPra1}, or leading to circular symmetry breaking and a complex dynamics of rotatory or random spots if the azimuthal instabilities dominate \cite{PORRASPra2}.

\begin{figure}[bh!]
\begin{center}
\includegraphics*[width=4.1cm]{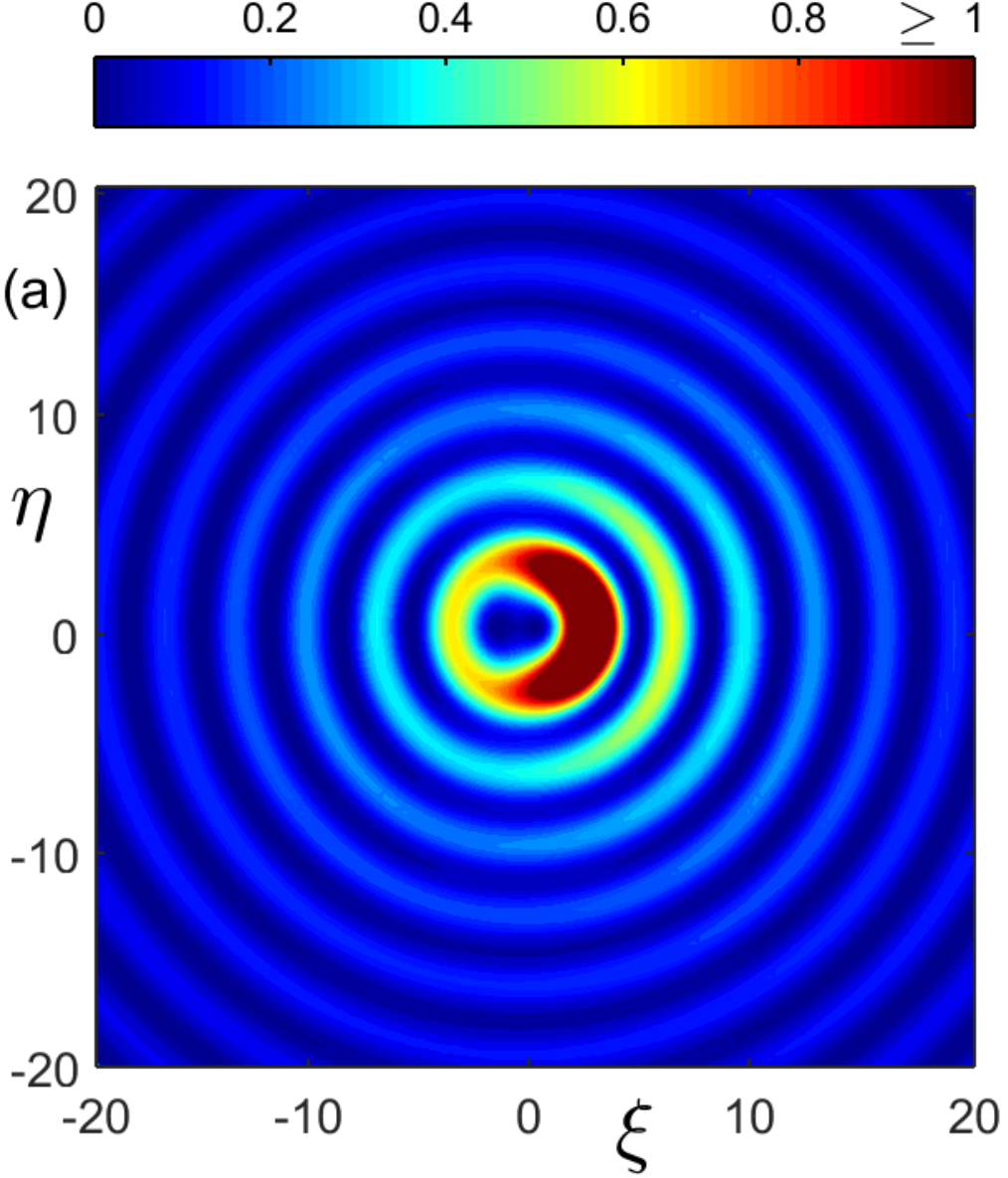}\includegraphics*[width=4.1cm]{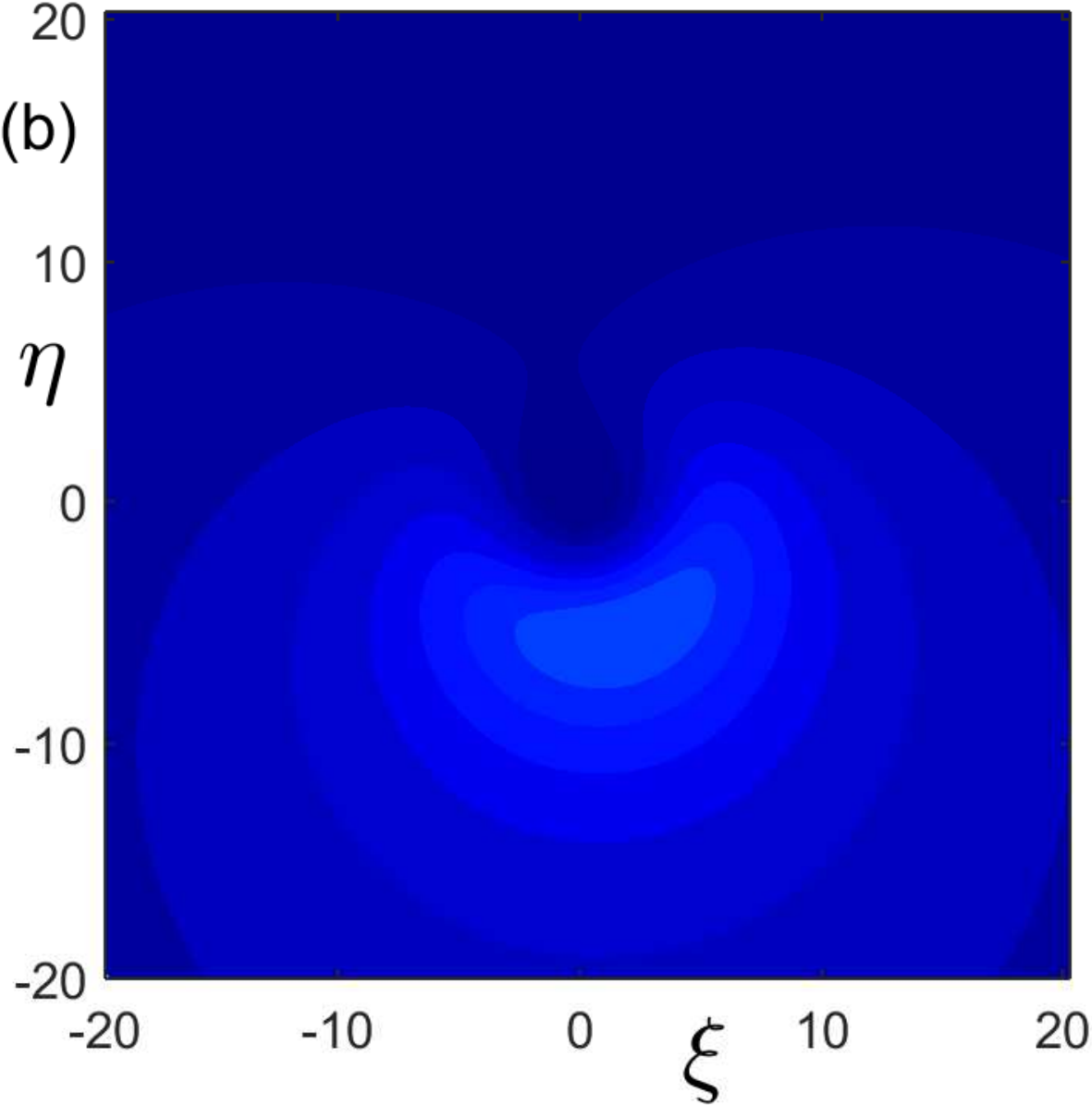}
\includegraphics[width=8.3cm]{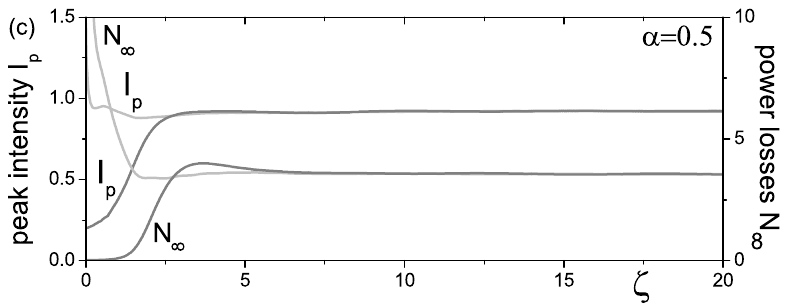}
\includegraphics*[width=4.1cm]{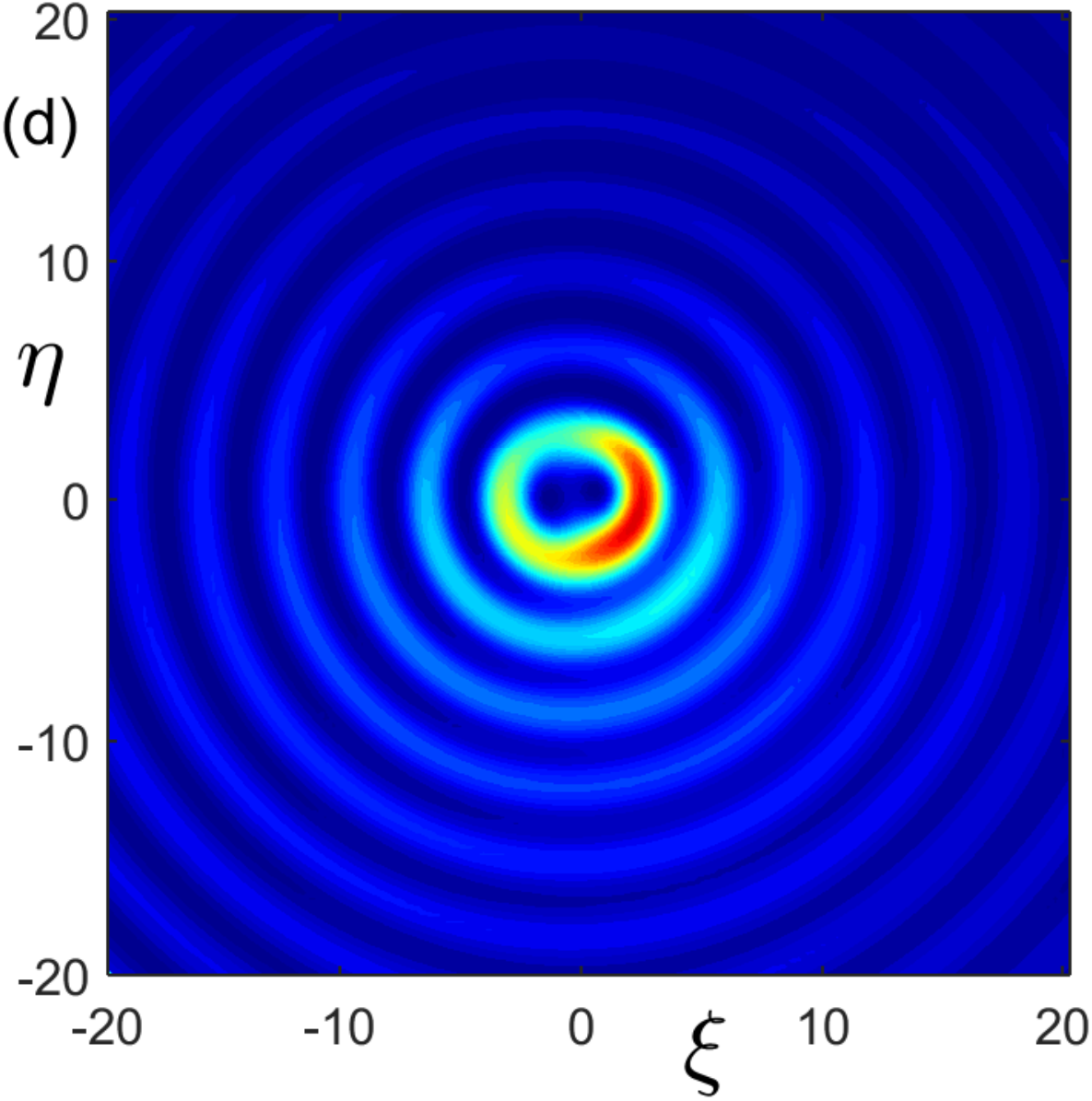}\includegraphics*[width=4.1cm]{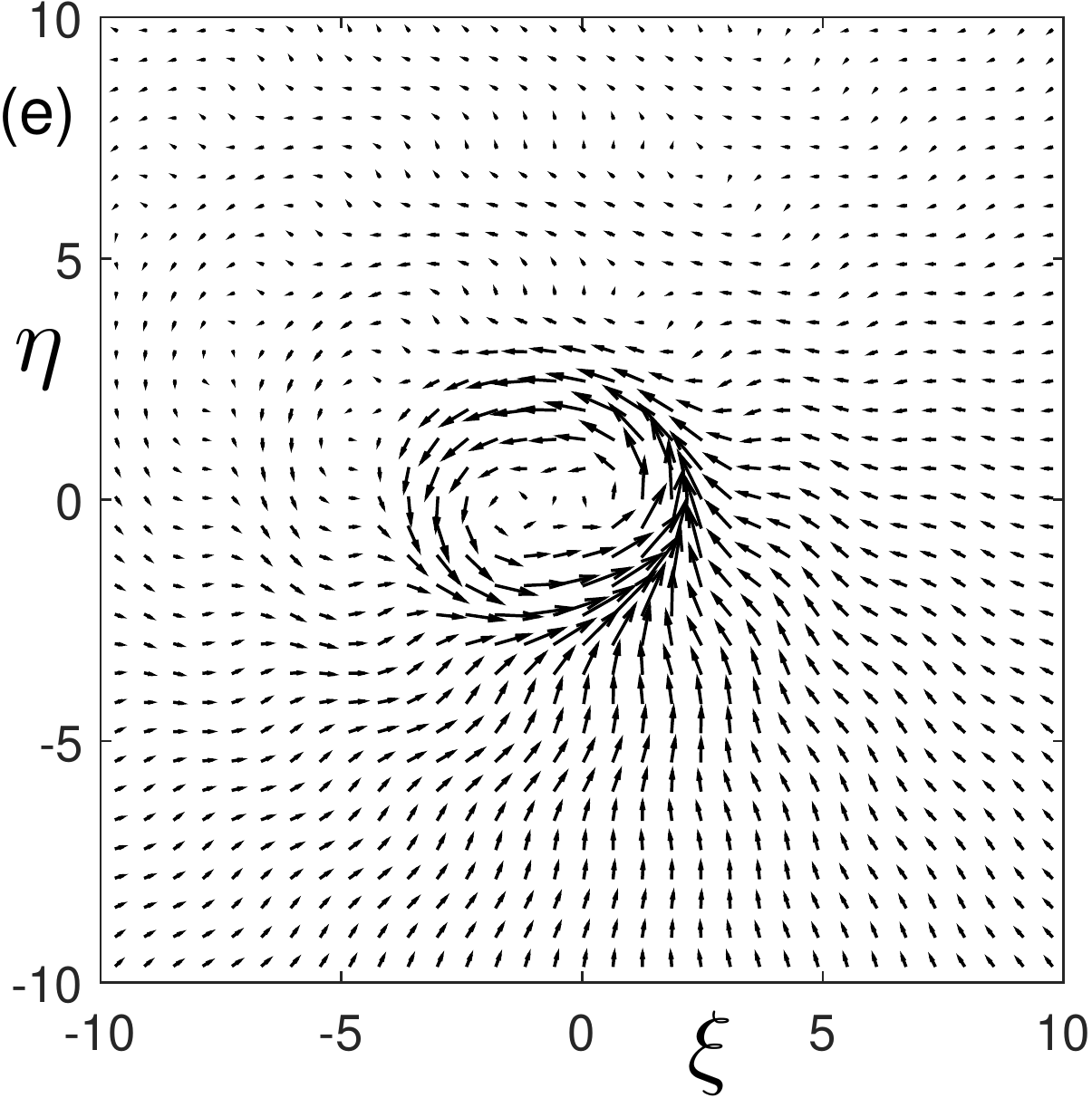}
\end{center}
\caption{\label{Fig2} Intensity of the input superposition of (a) two Bessel beams with $s_1=2$, $b_{L,1}=2$ and $s_2=1$, $b_{L,2}=1$, and (b) two inward H\"ankel beams with the same parameters multiplied by $(1-e^{-\rho^2/\rho_0^2})$, $\rho_0=4$, to eliminate the H\"ankel singularities at the origin without modifying the inward current at large radius. (c) In a medium and for cone angle such that $M=5$ and $\alpha=0.5$, normalized peak intensity and nonlinear power losses versus propagation distance for the input Bessel superposition (light gray curves) and for the input H\"ankel superposition (dark gray curves). (d) and (e) The propagation-invariant intensity and current patterns at large propagation distance ($\zeta=30$). All quantities are dimensionless, as explained in the text. Dimensionless cartesian coordinates are $(\xi,\eta)=k\theta(x,y)$.}
\end{figure}

\begin{figure}[bh!]
\begin{center}
\includegraphics[width=8.3cm]{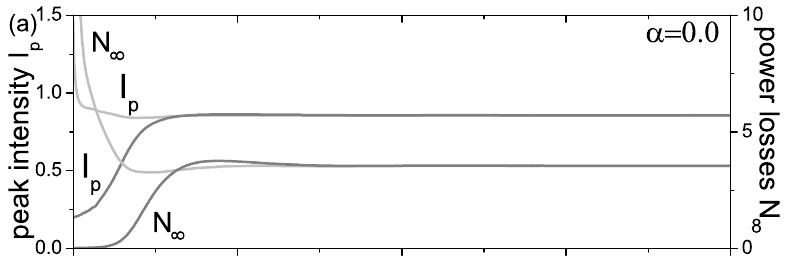}
\includegraphics[width=8.3cm]{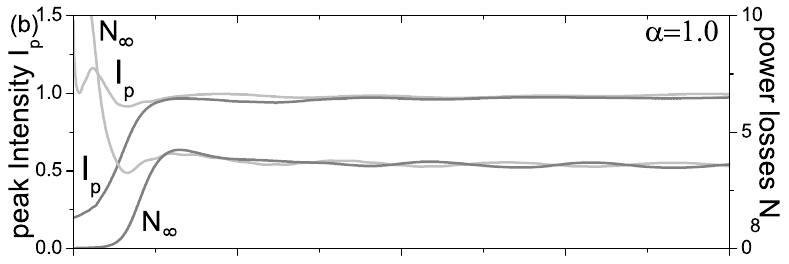}
\includegraphics[width=8.3cm]{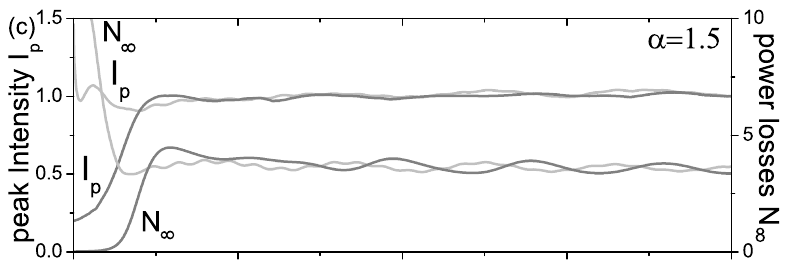}
\includegraphics[width=8.3cm]{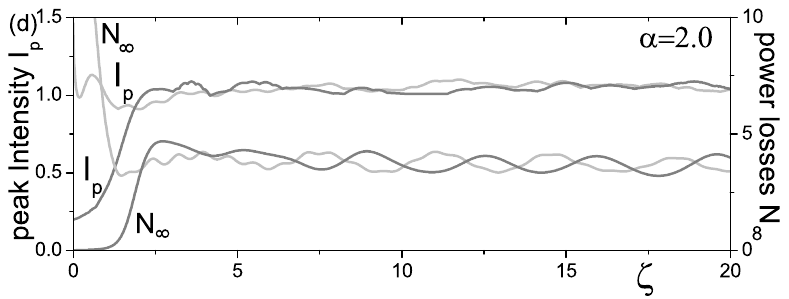}
\includegraphics*[width=4.1cm]{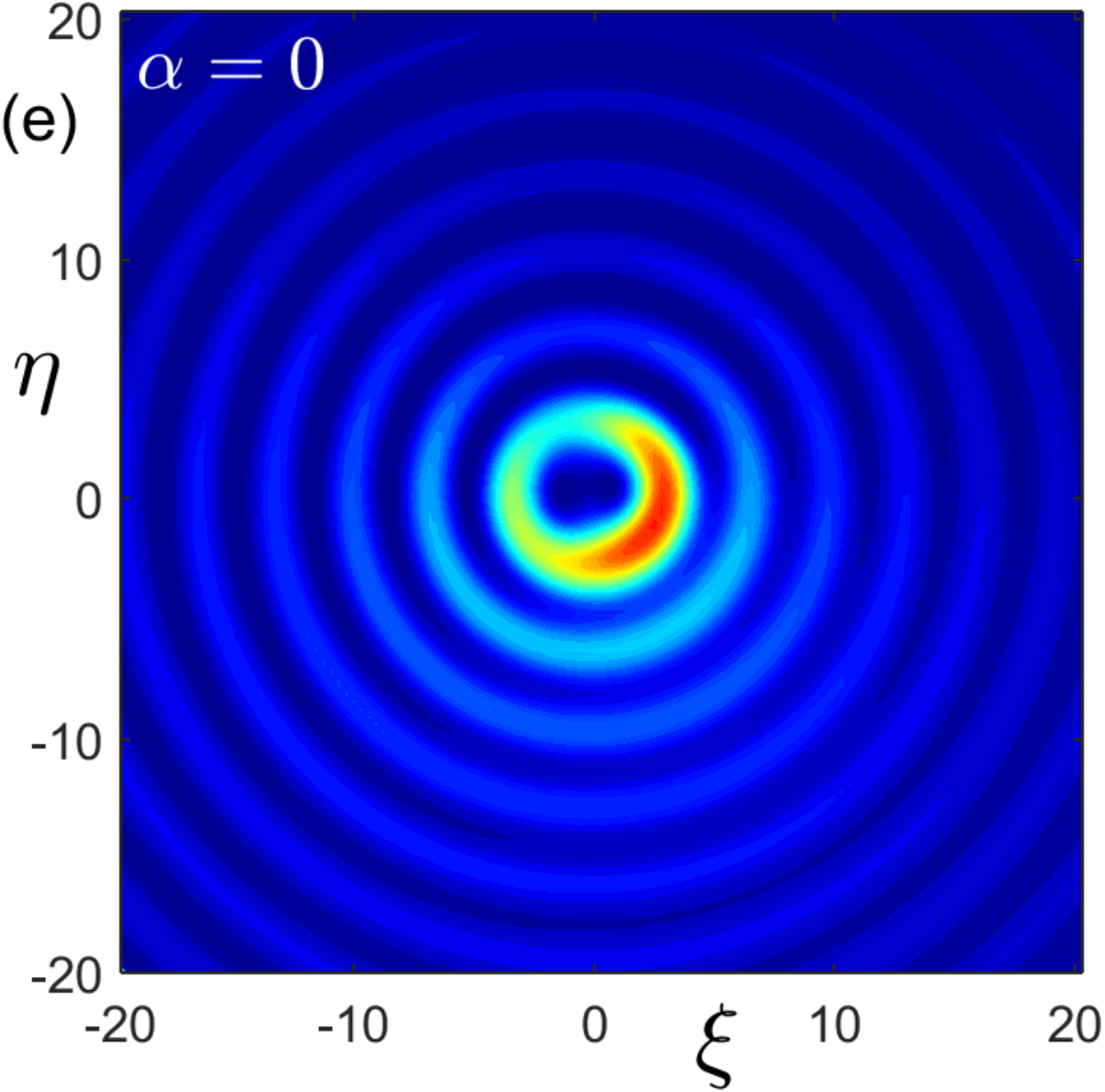}\includegraphics*[width=4.1cm]{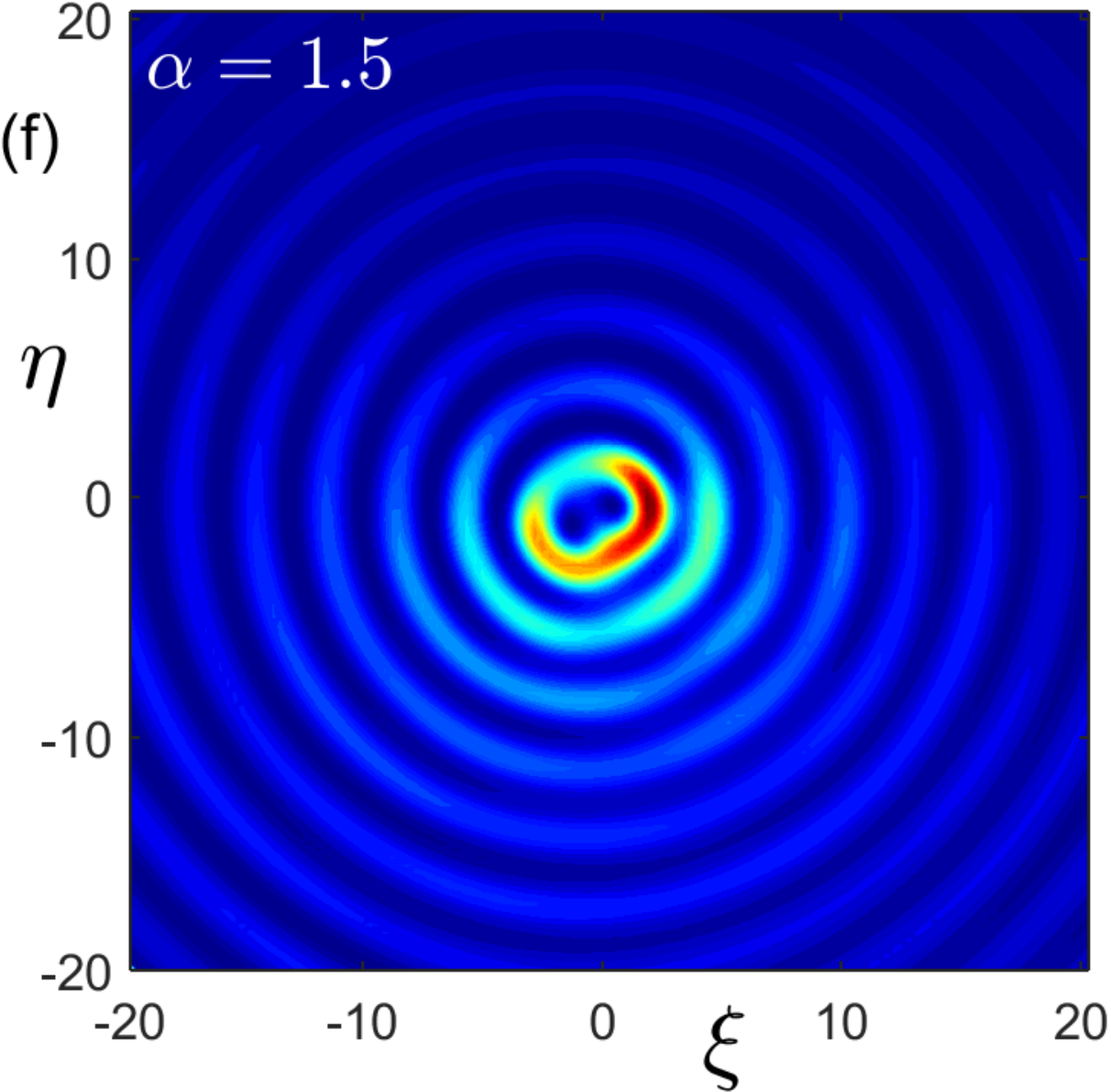}
\end{center}
\caption{\label{Fig3} (a-d) For the same input Bessel and H\"anhel superpositions as in Fig. \ref{Fig2}, peak intensity and nonlinear power losses versus propagation distance for $M=5$ and increasing values of $\alpha$. For the larger values of $\alpha$ these quantities do not reach constant values, indicating that a propagation-invariant state is not formed. (e) and (f) Intensity profiles of the final propagation-invariant states for  $\alpha\simeq 0.0$ and the maximum value $\alpha=\alpha_{\rm max}\simeq 1.5$ leading to the formation of a propagation-invariant beam. Their intensity profiles are increasingly compressed with increasing strength of the Kerr nonlinearity.}
\end{figure}

\section{Dissipatons}\label{DISSIPATONS}

With the above in mind, we investigate on the beam propagation dynamics that results from illuminating the medium with the coherent superposition $\tilde A(\rho,\varphi,\zeta=0)=\sum_{j} b_{L,j} J_{s_j}(\rho)e^{is_j\varphi}$ of several Bessel beams with same cone angle but different topological charges $s_j$ and amplitudes $b_{L,j}$. A large variety of input beam profiles can be obtained with different choices of the parameters of these superpositions, which are easily generated in laboratories \cite{FROEHLYJosaA,CLARKOe,VALILYEUOe}. In this section we consider the most general situation where this input beam does not have any particular symmetry in its transversal sections. As above, we will also consider illuminating the medium with other beams carrying the same inward intensity currents, as $\tilde A(\rho,\varphi,\zeta=0) = \sum_{j}\frac{1}{2} b_{L,j} H_{s_j}^{(2)}(\rho)e^{is_j\varphi} + f(\rho,\varphi)$ at large radius.

In Fig. \ref{Fig2} we have chosen two superposed beams with $s_1=2$, $b_{L,1}=2$ and $s_2=1$, $b_{L,2}=1$ in a medium and with cone angle such that $M=5$ and $\alpha=0.5$. The intensity patterns of the input Bessel beam superposition [Fig. \ref{Fig2}(a)] and of the input H\"ankel beam superposition [Fig. \ref{Fig2}(b)] differ substantially, so also their propagation in the medium. The input Bessel superposition is strongly absorbed at first, while input H\"ankel superposition builds up [Fig. \ref{Fig2}(c)]. With both input beams, however, the peak intensity and total power losses reach the same positive constant values at long distances. These values correspond to the same final, nonlinear stationary state that attracts the two beams, whose intensity and intensity current patterns in a transversal plane are shown in Figs. \ref{Fig2}(d) and (e).

These numerical simulations, and others below, show that the Kerr medium with nonlinear losses supports stationary propagation of beams that do not have any specific symmetry, to which we have referred in the introduction as ``dissipatons". In the example of Fig. \ref{Fig2}, the final state has even lost the bilateral symmetry of the input beam. The two vortices of unit charge in the final state are slightly displaced in the dissipaton from those in the input Bessel beam superposition.

In Figs \ref{Fig3}(a) to (d) we have plotted the peak intensity and nonlinear power power losses as functions of the propagation distance for the same couple of input Bessel and H\"ankel superpositions as above but several increasing values of $\alpha$ (e. g., increasing $n_2$, or diminishing the cone angle). Above a certain threshold, say $\alpha_{\rm max}$, ($\alpha_{\rm max}\simeq 1.5$ in this example) that depends on the particular input superposition and $M$, the dynamics does not lead to a stationary propagation regime, regardless of whether the Bessel or the H\"ankel superpositions are introduced in the medium. The propagation-invariant intensity profiles of the dissipatons from $\alpha\simeq 0$ (large enough cone angle or negligible Kerr nonlinearity) and to the maximum value $\alpha_{\rm max}\simeq 1.5$ [Figs. \ref{Fig3} (e) and (f)] feature increasing compression and irregularity of the nonlinear core.

\begin{figure}[t]
\begin{center}
\includegraphics*[width=4.3cm]{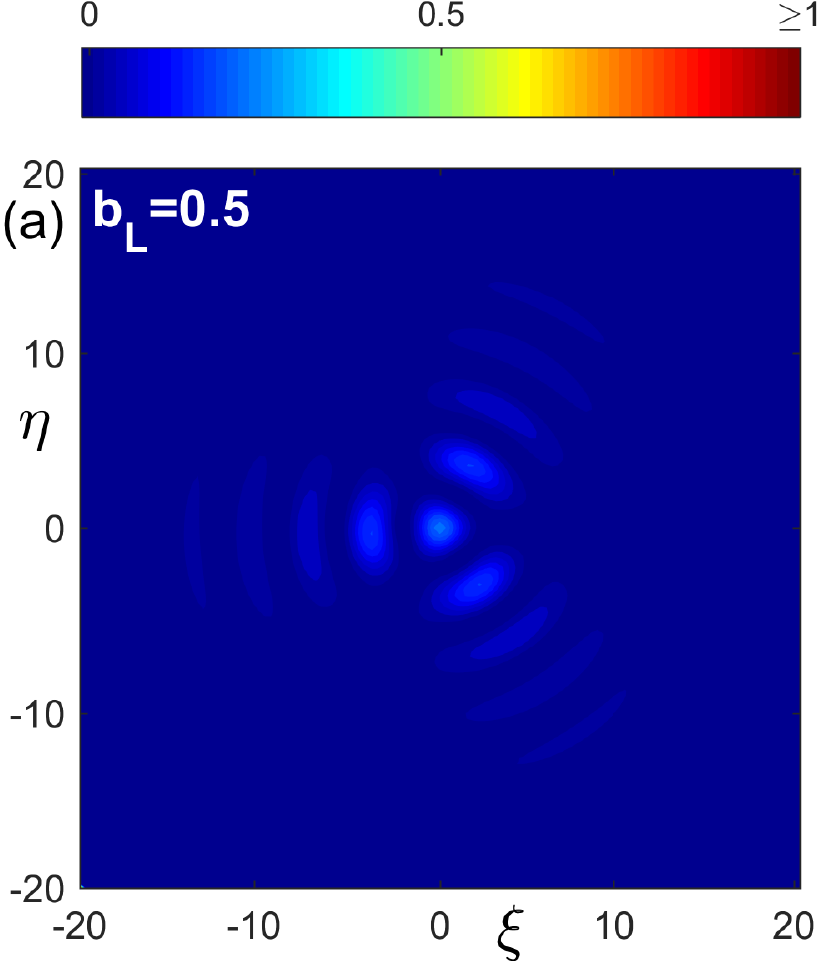}\includegraphics*[width=4.3cm]{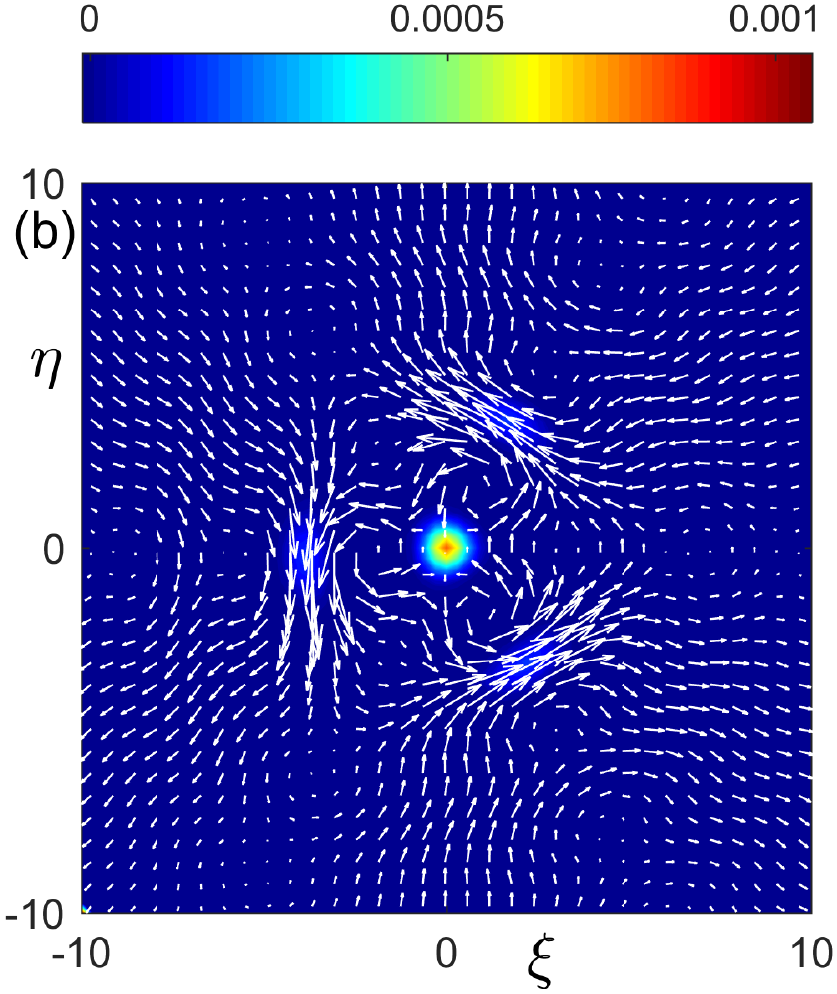}
\includegraphics*[width=4.3cm]{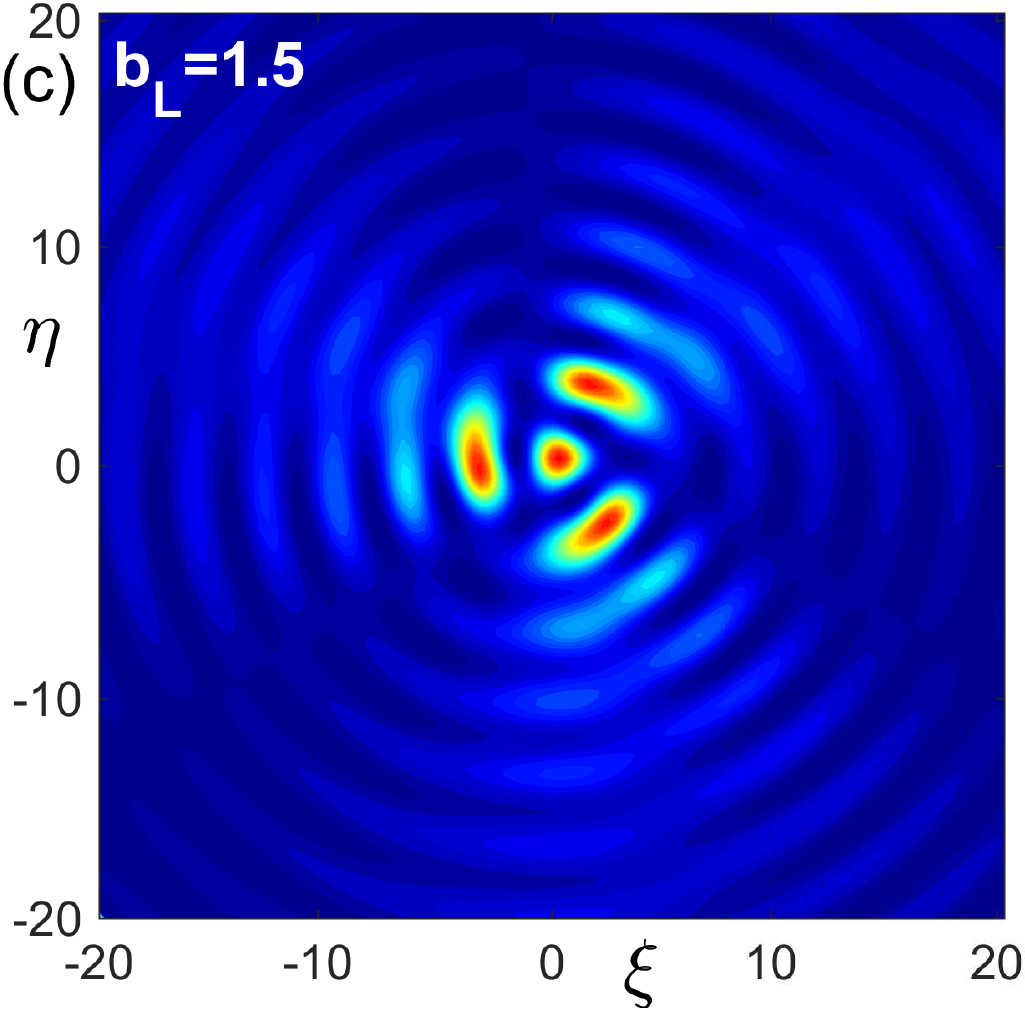}\includegraphics*[width=4.3cm]{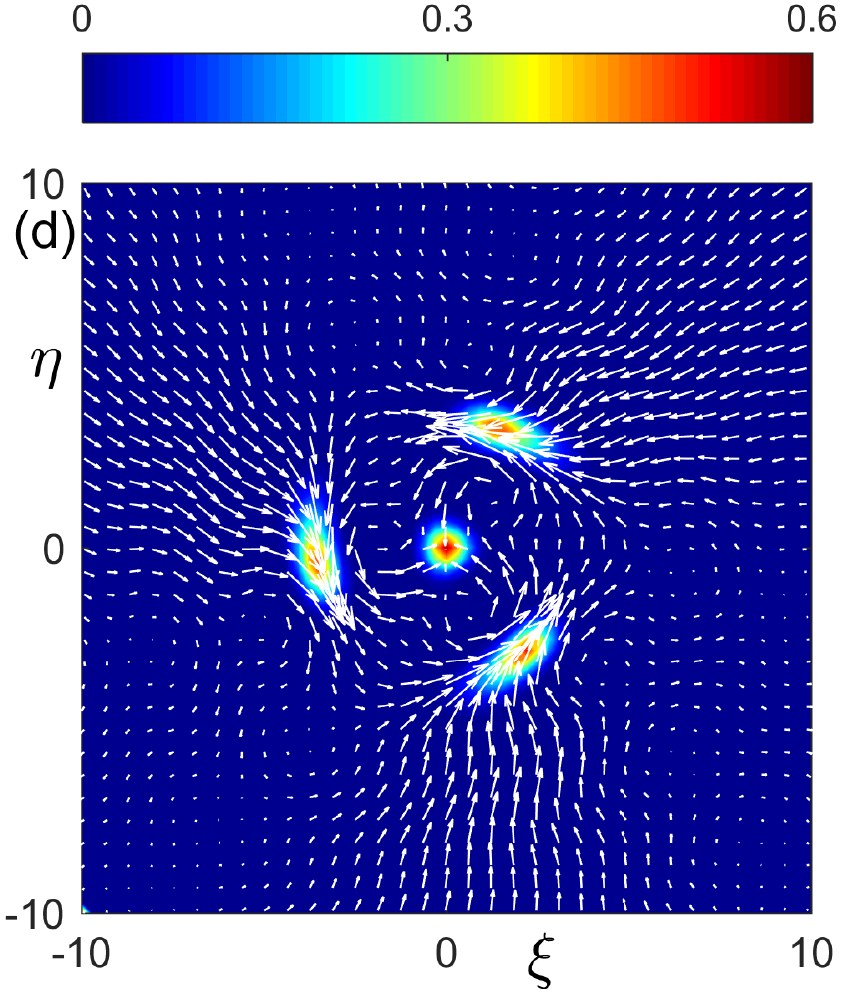}
\includegraphics*[width=4.3cm]{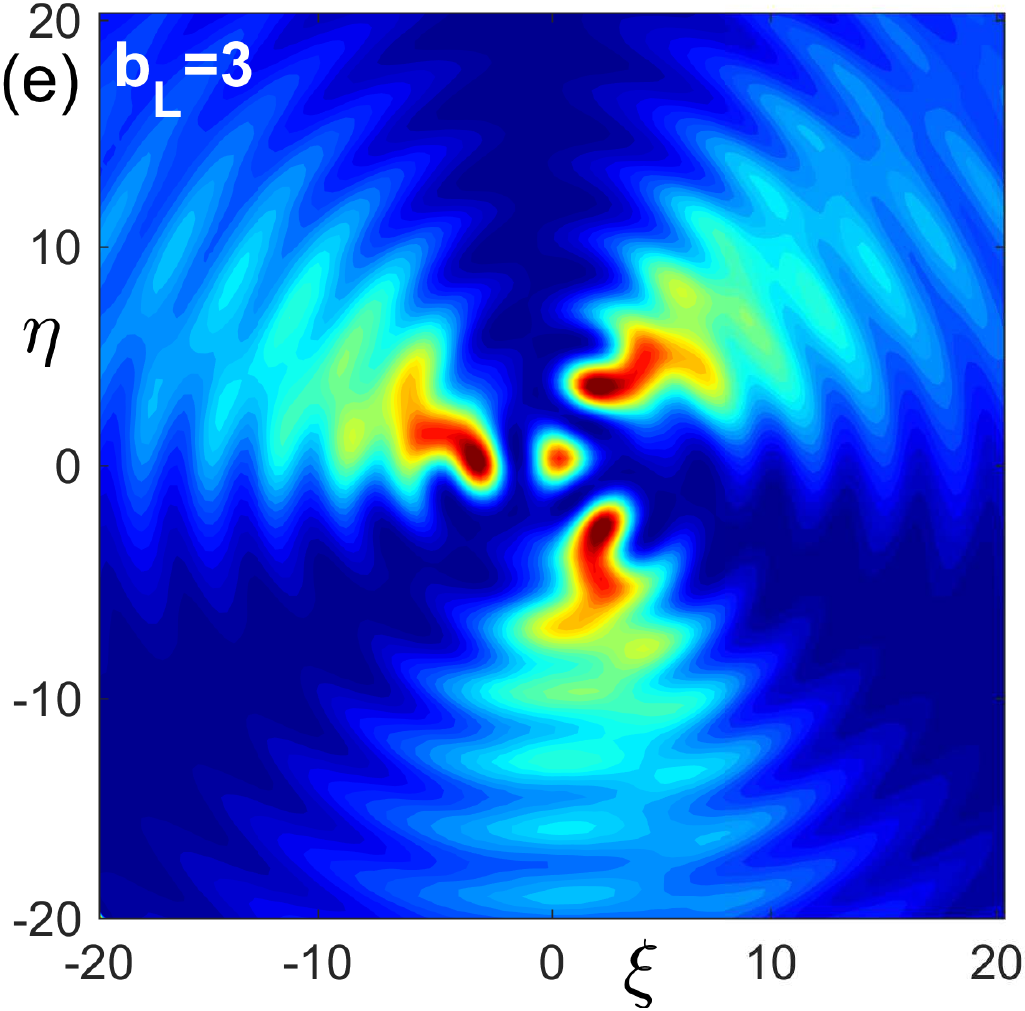}\includegraphics*[width=4.3cm]{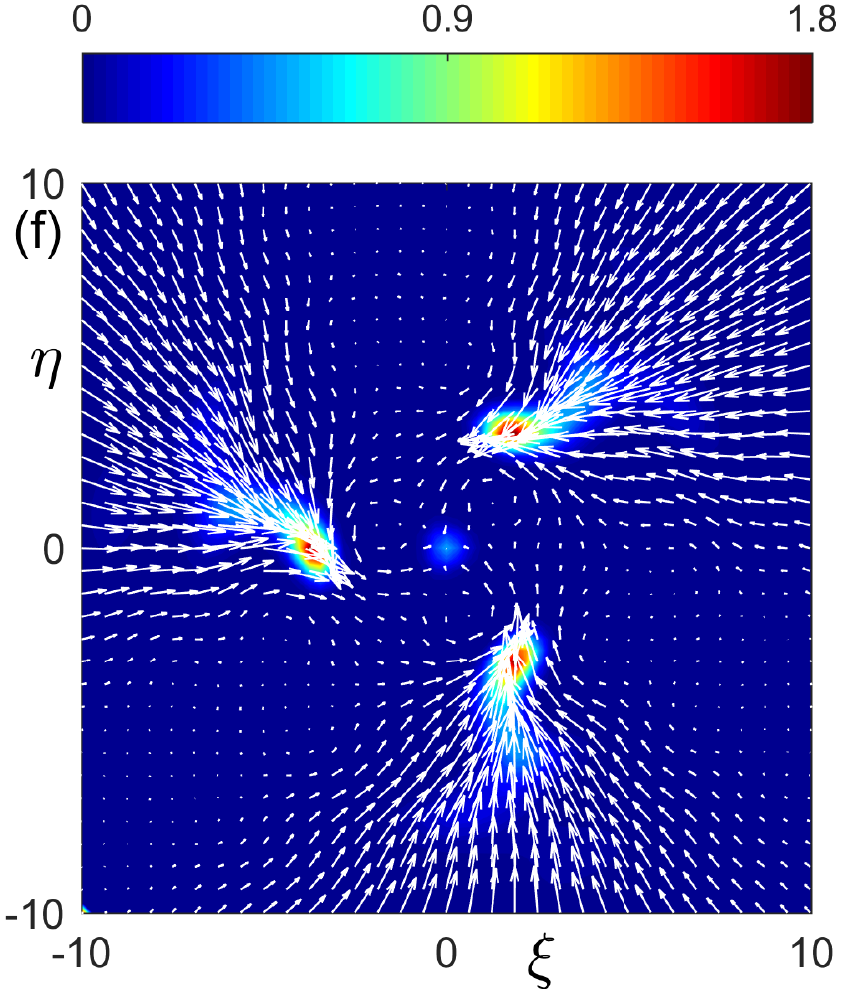}
\end{center}
\caption{\label{Fig4} Intensity patterns $|\tilde A|^2$ (left panels), and corresponding dissipation $|\tilde A|^{2M}$ and intensity current $\mathbf j$ patterns (right panels) of the dissipatons formed with cone angle and media such that $\alpha=0.5$ and $M=5$ starting from two conical beam superposition with $s_1=3$ and $s_2=0$, and increasing amplitudes $b_{L,1}=b_{L,2}=b_L=0.5, 1.5$ and $3$.}
\end{figure}

Similar results are seen to hold for other input beams. In the example of Fig. \ref{Fig4} we wanted to produce a triangular dissipaton with a a fourth hot spot in its center. For this we chose Bessel or H\"ankel superpositions with topological charges $s_1=3$ and $s_2=0$. All intensity patterns in Fig. \ref{Fig4} correspond to the same cone angle and the same medium ($\alpha=0.5$ and $M=5$), but increasing intensities  ($b_{L,1}=b_{L,2}\equiv b_L =0.5, 1.5$ and $3$). As above, these patterns become increasingly irregular, and no stationary state is reached above a certain value of $b_L$ that depends on the values of other parameters (above $b_L=3$ in this example). Figure \ref{Fig4} also shows the respective current patterns $\mathbf j$ and their associated sinks equal to the dissipation patterns $|\tilde A|^{2M}$. An input power such that $b_L=1.5$ produces four approximately equal dissipative spots. These examples, and others considered in Sec.  \ref{AZIMUTHONS}, illustrate that dissipation only occurs in strongly localized spots and not in the outer rings of the intensity pattern, and that these dissipative spots can be organized almost arbitrarily with an adequate choice of the topological charges and intensities of the input beam superposition.

\begin{figure}[b!]
\begin{center}
\includegraphics*[width=8.5cm]{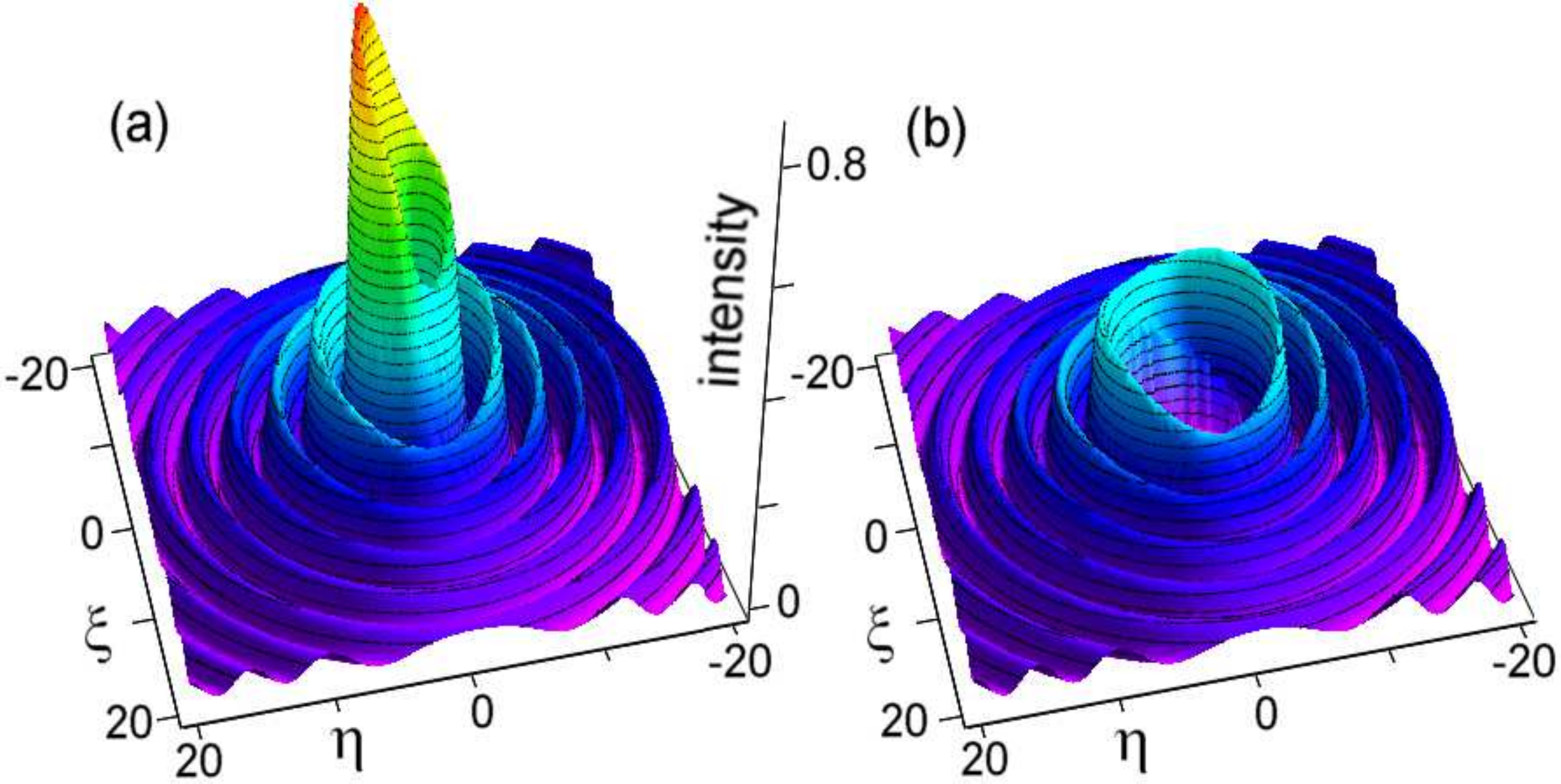}
\includegraphics*[width=8cm]{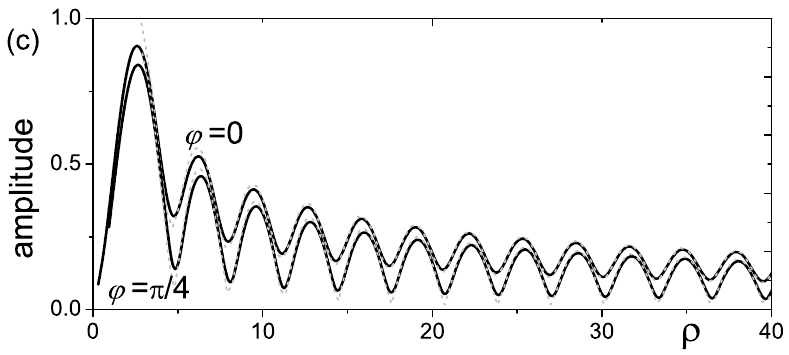}\\
\includegraphics*[width=4cm]{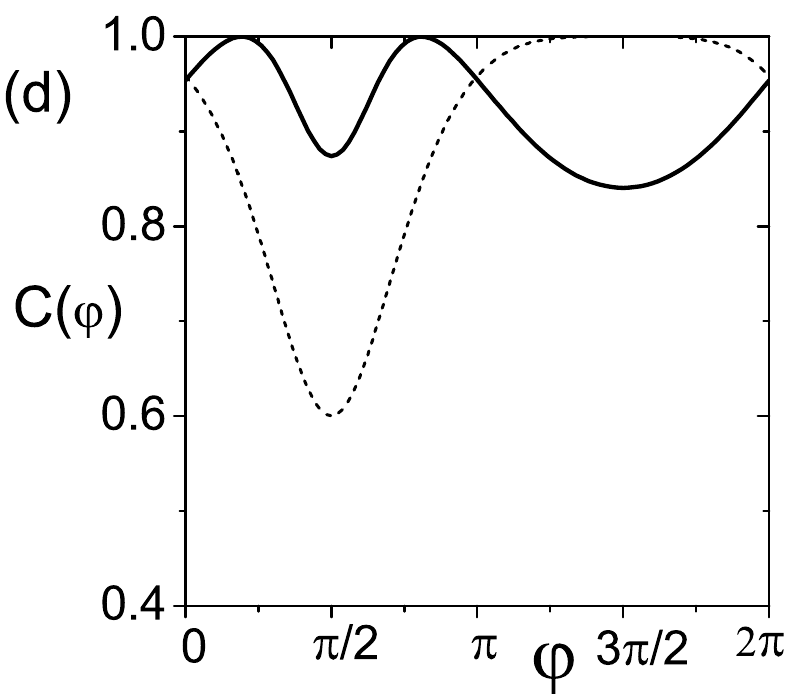}\includegraphics*[width=4cm]{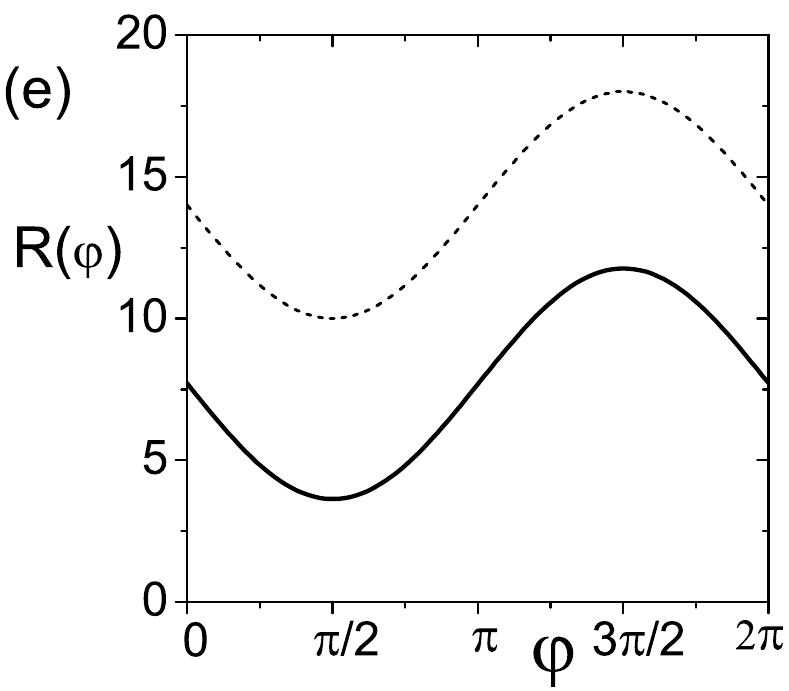}
\end{center}
\caption{\label{Fig5} (a) Intensity distribution of the dissipaton formed in media with $M=5$ and $\alpha\simeq 0$ starting with the Bessel beam superposition with $s_1=2$, $b_{L,1}=2$ and $s_2=1$, $b_{L,2}=1$. (b) Intensity distribution (for $\rho>5$) of the asymptotic form in Eq. (\ref{ASYMP1}) with $b_{{\rm in},1}=b_{L,1}=2$, $b_{{\rm in},2}=b_{L,2}=1$, $b_{{\rm out},1}\simeq 1.11$ and $b_{{\rm out},2}=0.64\,e^{i 0.70}$, verifying $|b_{{\rm in},1}|^2+|b_{{\rm in},2}|^2-|b_{{\rm out},1}|^2-|b_{{\rm out},2}|^2=N_\infty=3.35$ [see Fig. \ref{Fig3}(e)]. (c) Radial amplitude distributions (black solid curves) and their asymptotic fitting with the above values (gray dashed curves) for the azimuthal angles $\varphi=0$ and $\varphi=\pi/4$, featuring decaying oscillations of different contrasts about different mean values. The amplitude instead of the intensity is shown for a better visibility. (d) and (e) Contrast and average value of the radial oscillations at each azimuthal direction obtained from the above values of the H\"ankel beam amplitudes and Eqs. (\ref{CPHI}) and (\ref{RPHI}). The dashed curves are the same quantities for the input Bessel beam superposition.}
\end{figure}

\subsection*{Asymptotic behavior of dissipatons}

The structure of the non-dissipative outer part of the dissipatons is found to be simpler than that of the dissipative spots. At fixed azimuthal angle $\varphi$ and increasing radius the intensity is always seen to decay as $\rho^{-1}$ while oscillating harmonically. The same happens with the circularly symmetric nonlinear BVBs \cite{PORRASPrl,PORRASJosaB}, but now the decay rate and the oscillation contrast depend on the specific azimuthal angle. This leads us to conjecture a linear conical structure for these tails of the form
\begin{equation}\label{ASYMP1}
\tilde A\simeq \sum_{j} \frac{1}{2}\left[b_{{\rm out},j}H_{s_j}^{(1)}(\rho)+ b_{{\rm in},j}H_{s_j}^{(2)}(\rho)\right]e^{is_j\varphi}e^{-i\zeta}\, ,
\end{equation}
i. e., a linear superposition of unbalanced Bessel beams. Using Eq. (\ref{HANKEL}) for H\"ankel beams at large radius, Eq. (\ref{ASYMP1}) can also be written as
\begin{equation}\label{ASYMP2}
\tilde A\simeq \frac{1}{\sqrt{2\pi\rho}}\left[b_{\rm out}(\varphi)e^{i(\rho-\pi/4)}+ b_{\rm in}(\varphi)e^{-i(\rho-\pi/4)}\right]e^{-i\zeta} \, ,
\end{equation}
where we have defined
\begin{eqnarray}\label{BPHI}
b_{\rm out}(\varphi) & \equiv &\sum_{j} b_{{\rm out},j} e^{is_j(\varphi-\pi/2)}\,, \\
b_{\rm in} (\varphi) & \equiv &\sum_{j} b_{{\rm in},j} e^{is_j(\varphi+\pi/2)}\, .
\end{eqnarray}
Equation (\ref{ASYMP2}) then yields an intensity profile at large radius given by
\begin{eqnarray}\label{RADIALI}
2\pi\rho|\tilde A|^2\!&\simeq&\! |b_{\rm out}(\varphi)|^2\!+\!|b_{\rm in}(\varphi)|^2\!+\!2|b_{\rm out}(\varphi)||b_{\rm in}(\varphi)|  \nonumber\\
     &\times &
\cos\left[2(\rho-\pi/4)\! +\kappa(\varphi)\right]\,,
\end{eqnarray}
where $\kappa(\varphi)=\mbox{arg}b_{\rm out}(\varphi)-\mbox{arg}b_{\rm in}(\varphi)$. As a function of $\rho$, Eq. (\ref{RADIALI}) represents  harmonic oscillations of $\varphi$-dependent contrast
\begin{equation}\label{CPHI}
C(\varphi)= \frac{2|b_{\rm out}(\varphi)| |b_{\rm in}(\varphi)|}{|b_{\rm out}(\varphi)|^2 + |b_{\rm in}(\varphi)|^2}
\end{equation}
about the $\varphi$-dependent mean value
\begin{equation}\label{RPHI}
R(\varphi)=|b_{\rm out}(\varphi)|^2 + |b_{\rm in}(\varphi)|^2 \, .
\end{equation}

The linearity of these tails allows us to generalize the result in the circularly symmetric case \cite{PORRASJosaB} that the cone angle, all topological charges $s_j$ and amplitudes of the inward H\"ankel beams in the asymptotic expression (\ref{ASYMP1}) of a dissipaton are the same as those of the input beam, i. e.,
\begin{equation}\label{CONS}
|b_{{\rm in},j}|=|b_{L,j}| \quad \mbox{for all $j$.}
\end{equation}
Also, the asymptotic expression for the radial component of the intensity current can be evaluated to be $j_\rho \simeq \mbox{Im}\left\{\tilde A^\star \partial_\rho \tilde A\right\}=\left[|b_{\rm out}(\varphi)|^2 - |b_{\rm in}(\varphi)|^2\right]/(2\pi\rho)$. From Eq. (\ref{CR}), we obtain the constant inward radial flux at large radius offsetting the total power losses as
\begin{equation}
-F_\infty =-\int_0^{2\pi}\frac{|b_{\rm out}(\varphi)|^2 - |b_{\rm in}(\varphi)|^2}{2\pi}d\varphi = N_\infty\, ,
\end{equation}
and upon integration, the relation
\begin{equation}\label{INFTY}
\sum_{j} |b_{{\rm in},j}|^2 - \sum_{j} |b_{{\rm out},j}|^2 = N_{\infty}
\end{equation}
between the amplitudes of the inward and outward H\"ankel beam components.

The validity of all previous relations is supported by the numerical analysis. As an example, Figs. \ref{Fig5} (a) and (b) show the intensity profile of a dissipaton [the same as in Figs. \ref{Fig3} (a) and (e)], and the fitted intensity profile (for $\rho>5$) using Eq. (\ref{ASYMP1}) with $b_{{\rm in},j}$ and $b_{{\rm out},j}$ satisfying Eqs. (\ref{CONS}) and (\ref{INFTY}) (see caption for detailed values). The radial profiles of the dissipaton and of the asymptotic form at two particular azimuthal angles are more clearly seen to match better and better with increasing radius in Fig. \ref{Fig5} (c). With the values of $b_{{\rm in},j}$ and $b_{{\rm out},j}$, one can evaluate $b_{\rm in}(\varphi)$ and $b_{\rm out}(\varphi)$ from Eqs. (\ref{BPHI}), and the contrast $C(\varphi)$ and average oscillation value $R(\varphi)$ of the radial oscillations at each azimuthal direction from Eqs. (\ref{CPHI}) and (\ref{RPHI}). These are plotted in Figs. \ref{Fig5}(d) and (e) in the above example. Compared to the same quantities for the input Bessel beam superposition (dashed curves), the average value of the oscillations is significantly smaller due to the initial stage of strong absorption. It is also characteristics of dissipatons that the contrast of the oscillations is significantly diminished along the azimuthal directions of high intensity, i. e., high $R(\varphi)$, as for the angle $\varphi=3\pi/2$ in this example.

Additional confirmation of the generality of the above description is given in Sec. \ref{AZIMUTHONS} for dissipatons with $n$-fold rotational symmetry. By analogy with nonlinear BVBs, each dissipaton is determined by the set of parameters $M$, $\alpha$, $s_j$ and $b_{L,j}$. We can conjecture, following the same analogy, that there exist dissipatons with any value of these parameters, as for nonlinear BVBs \cite{PORRASJosaB}, and that they may be stable or unstable for different choices of the parameters, e. g., unstable above $\alpha_{\rm max}$ once the other parameters are fixed. This hypothesis is suggested by the fact that when the propagation does not lead to a stationary state, as in Fig. \ref{Fig3}(d), the beam is not completely dissipated or dispersed, but its peak intensity, power losses, and the entire beam profile oscillates more or less randomly about constant values, as in the circularly symmetric case \cite{PORRASPra2}.

\section{Dissipatons with $2s$-fold rotational symmetry}\label{AZIMUTHONS}

\begin{figure}
\begin{center}
\includegraphics*[width=4.3cm]{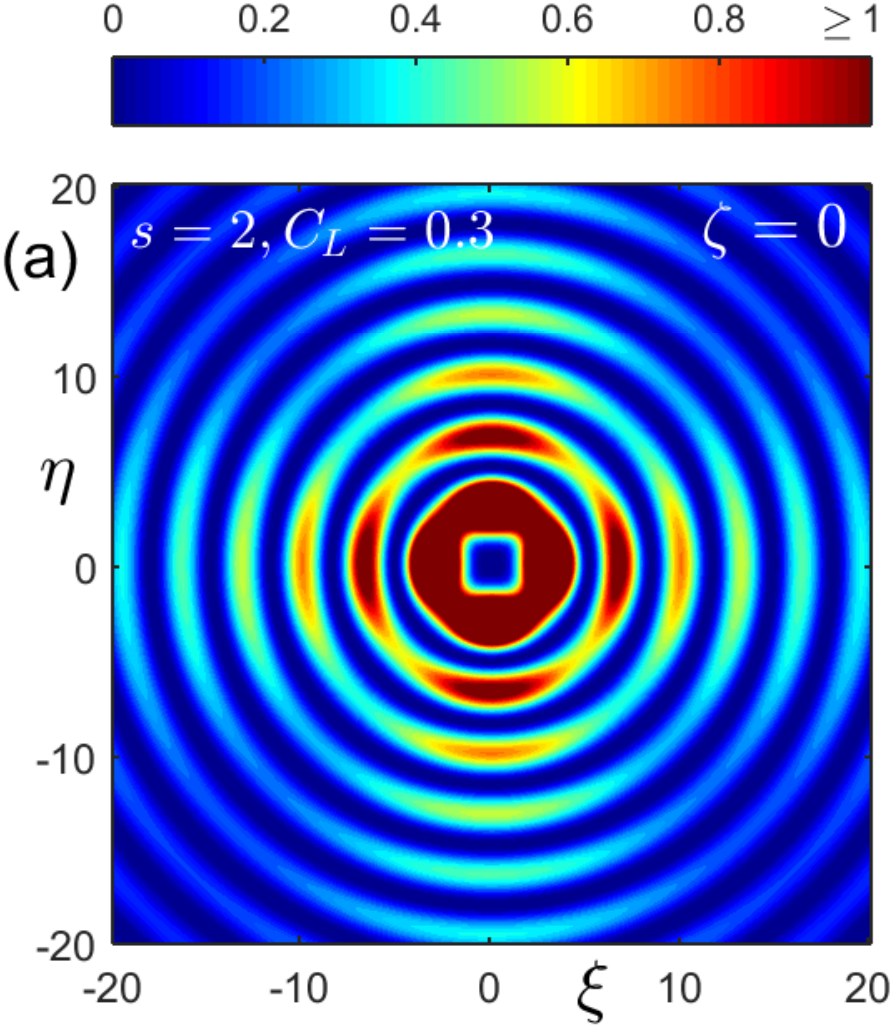}\includegraphics*[width=4.3cm]{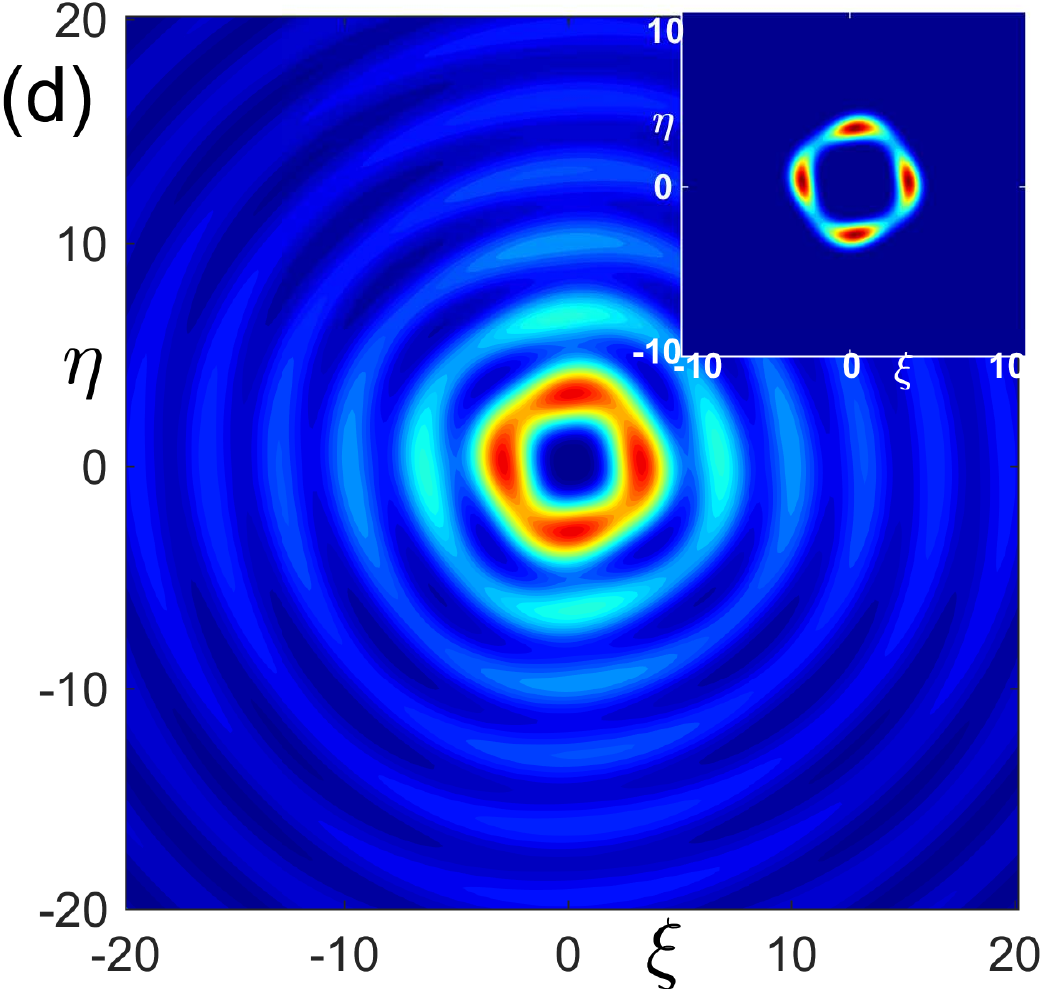}
\includegraphics*[width=4.3cm]{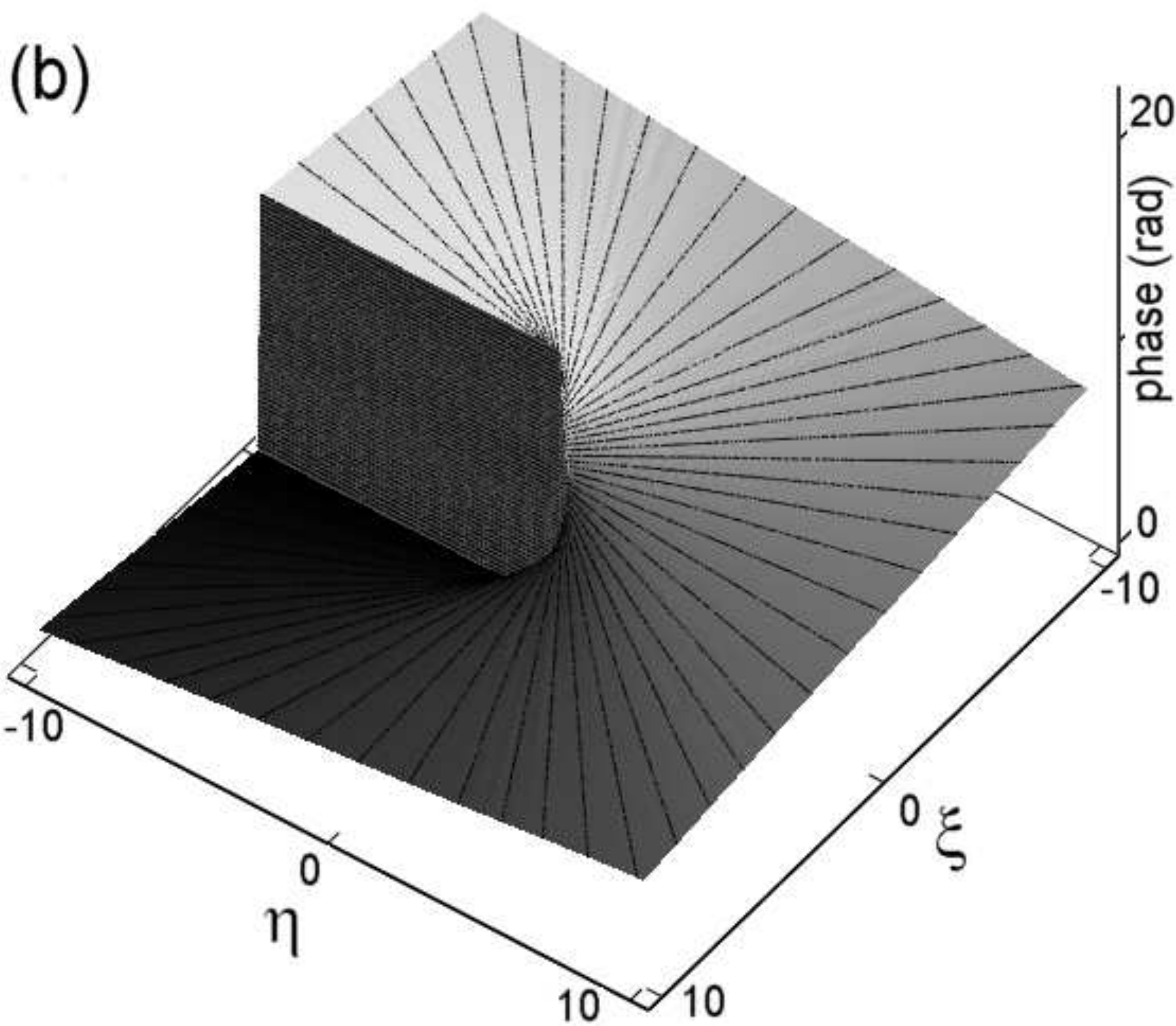}\includegraphics*[width=4.3cm]{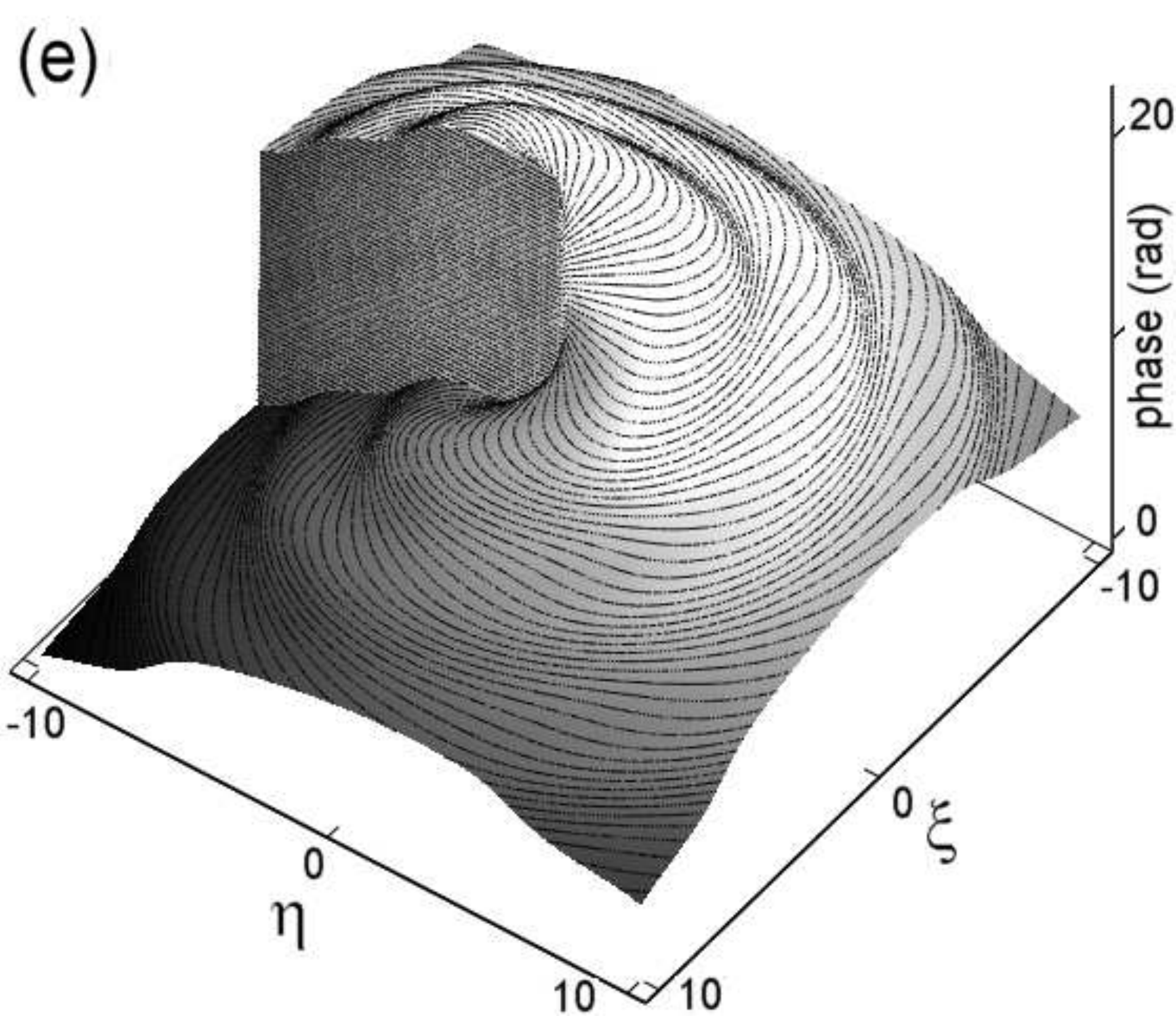}
\includegraphics*[width=4.3cm]{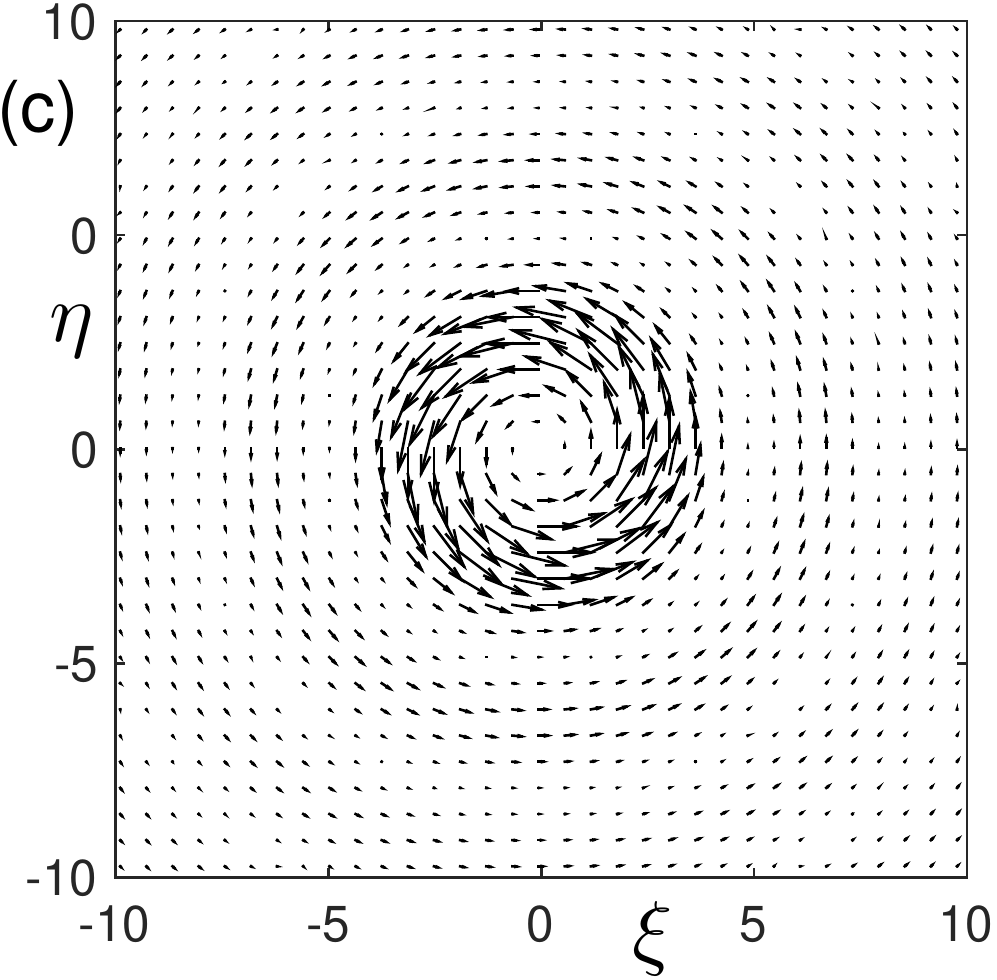}\includegraphics*[width=4.3cm]{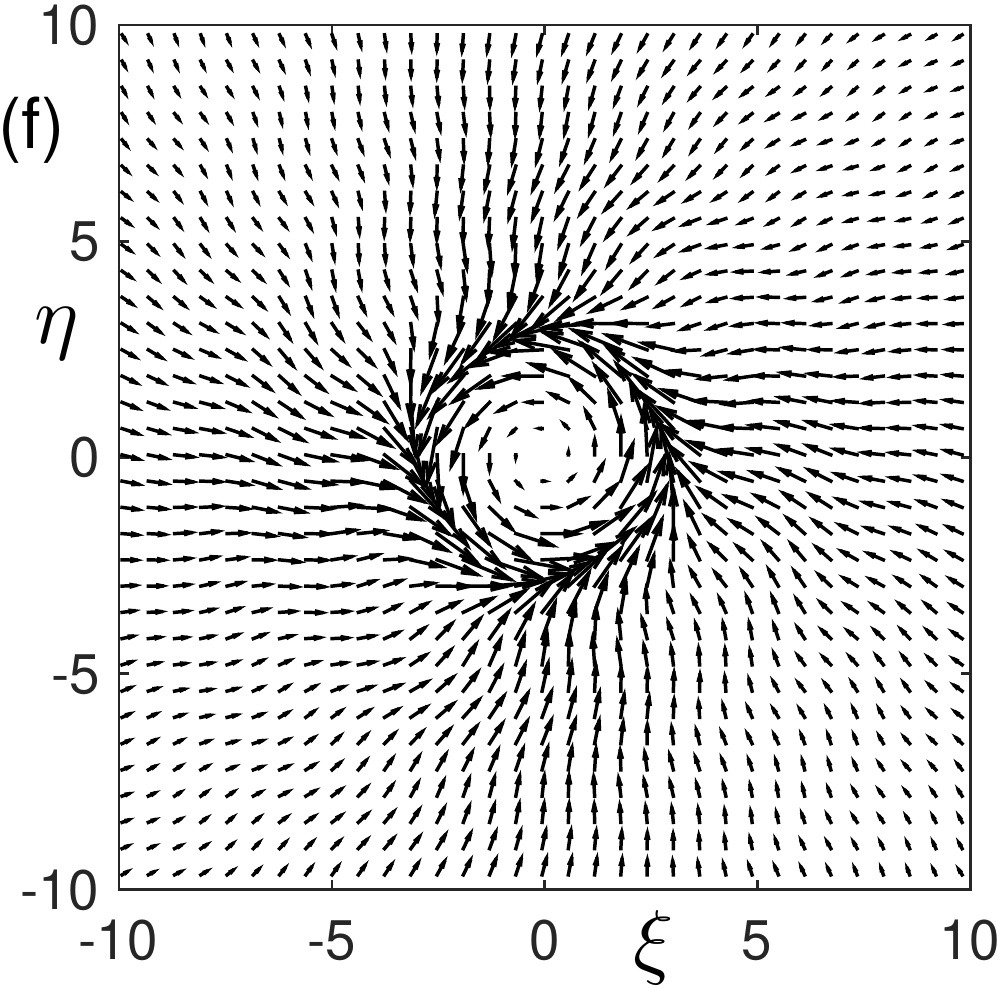}
\end{center}
\caption{\label{Fig6} (a) Intensity, (b) phase and (c) intensity current of the initial superposition of two Bessel beams of opposite charges, $s=2$ and $-s=-2$, average amplitude $b_L=3$ and azimuthal contrast $C_L=0.3$ ($b_{L,s}=2.965$ and $b_{L,-s}=0.455$). (d) Intensity, (e) phase and (f) intensity current of the final stationary state in the form of a dissipaton formed in a medium with $M=5$ and $\alpha\simeq 0$. The inset in (d) is the dissipation pattern $\tilde a^{2M}$.}
\end{figure}

\begin{figure}
\begin{center}
\includegraphics*[width=4.3cm]{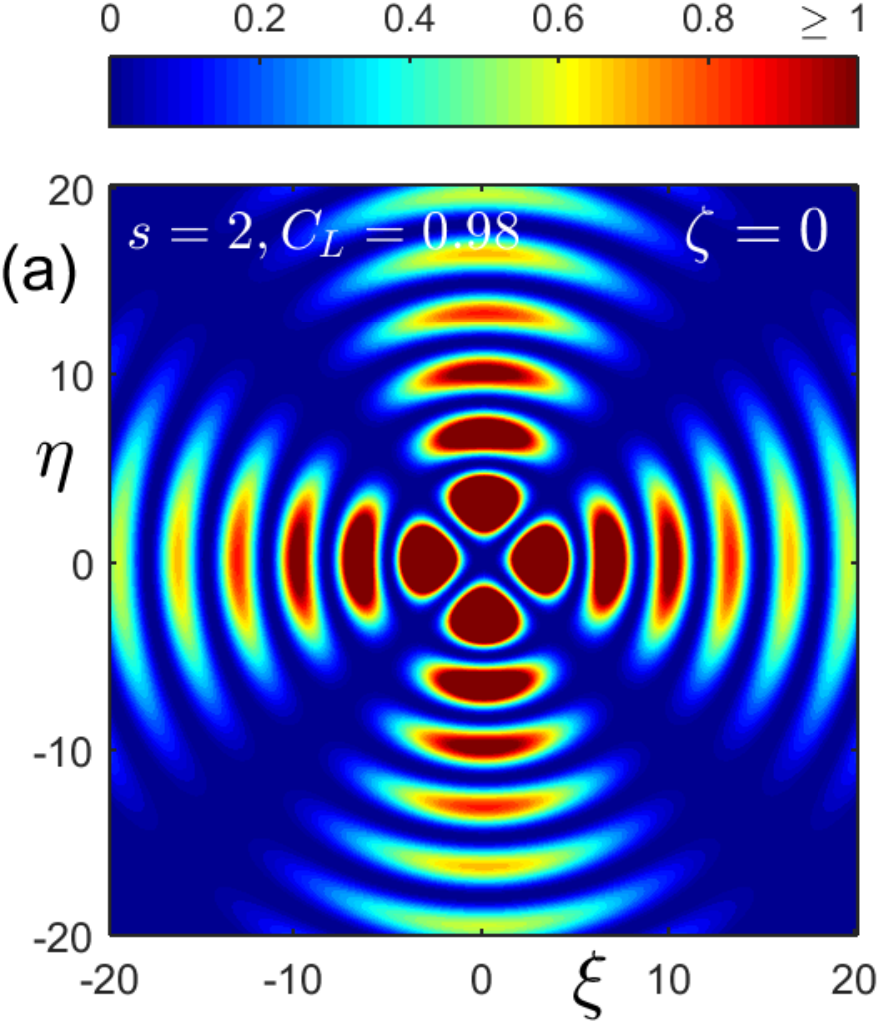}\includegraphics*[width=4.3cm]{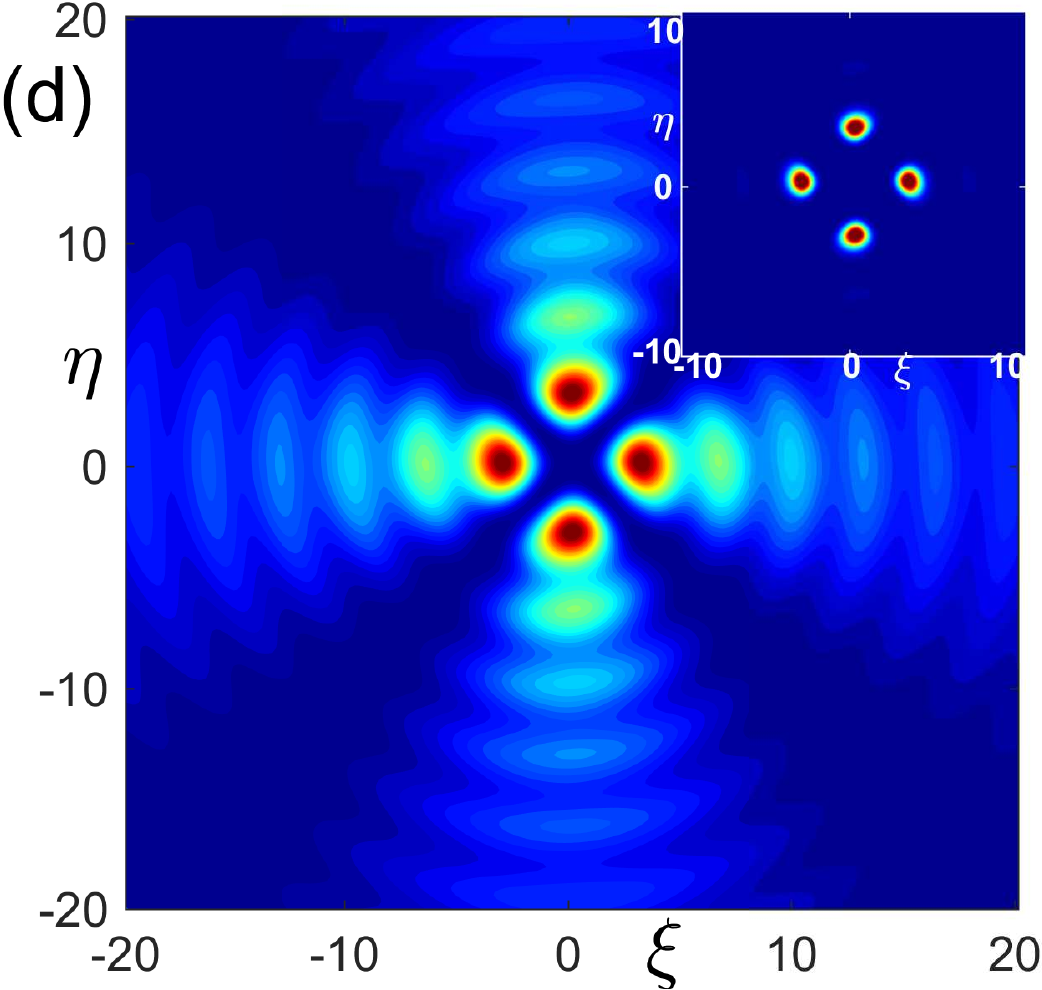}
\includegraphics*[width=4.3cm]{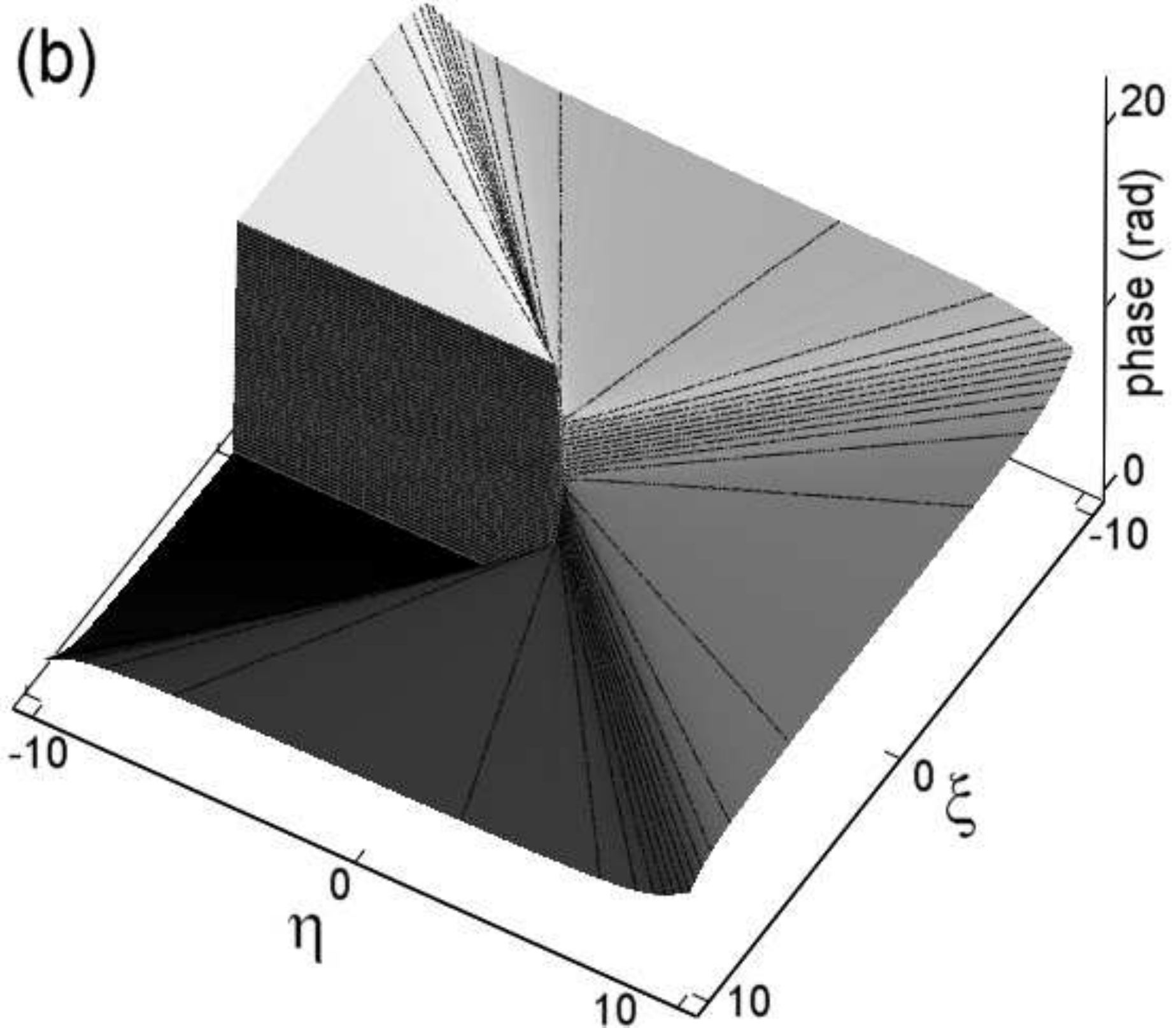}\includegraphics*[width=4.3cm]{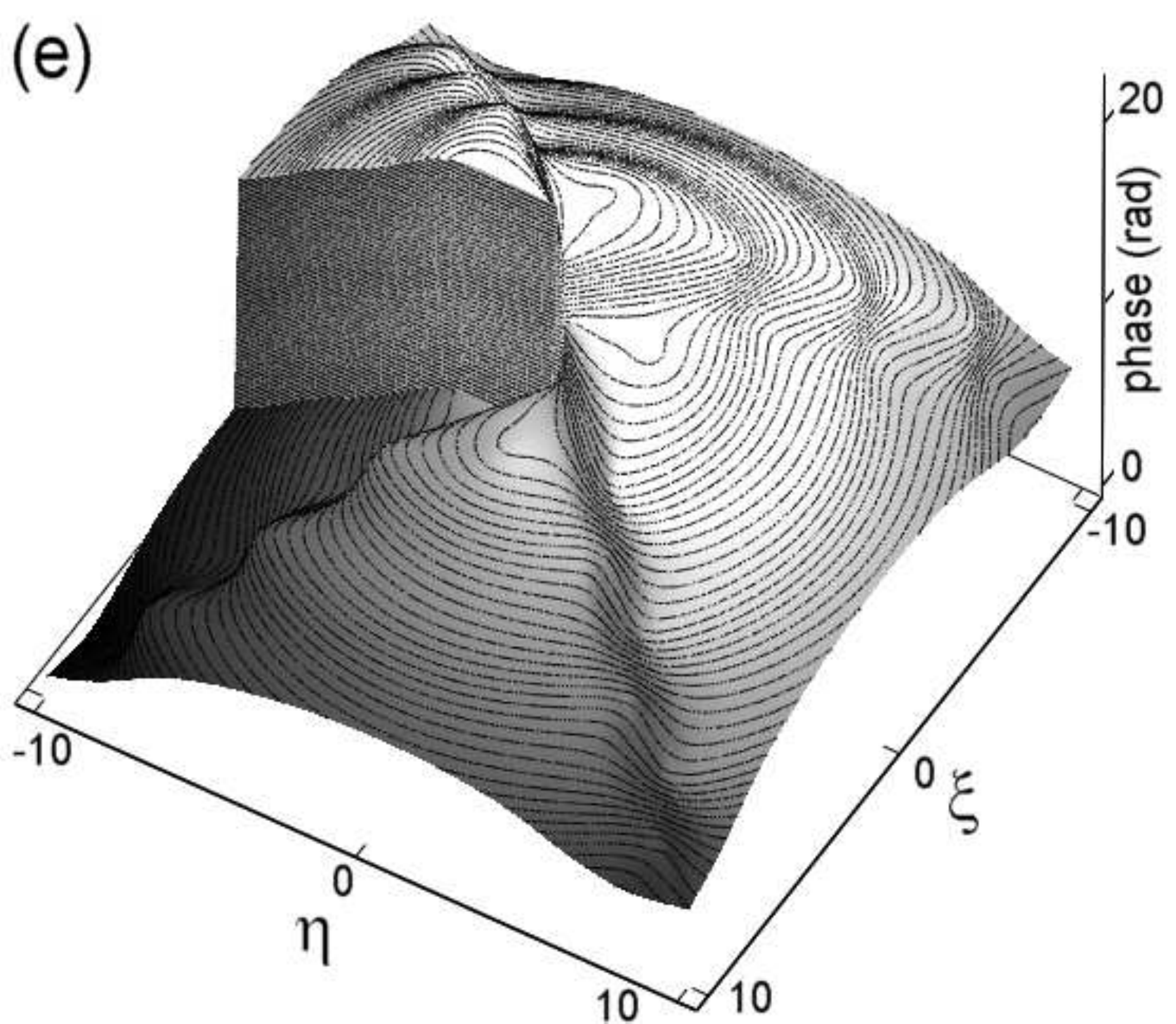}
\includegraphics*[width=4.3cm]{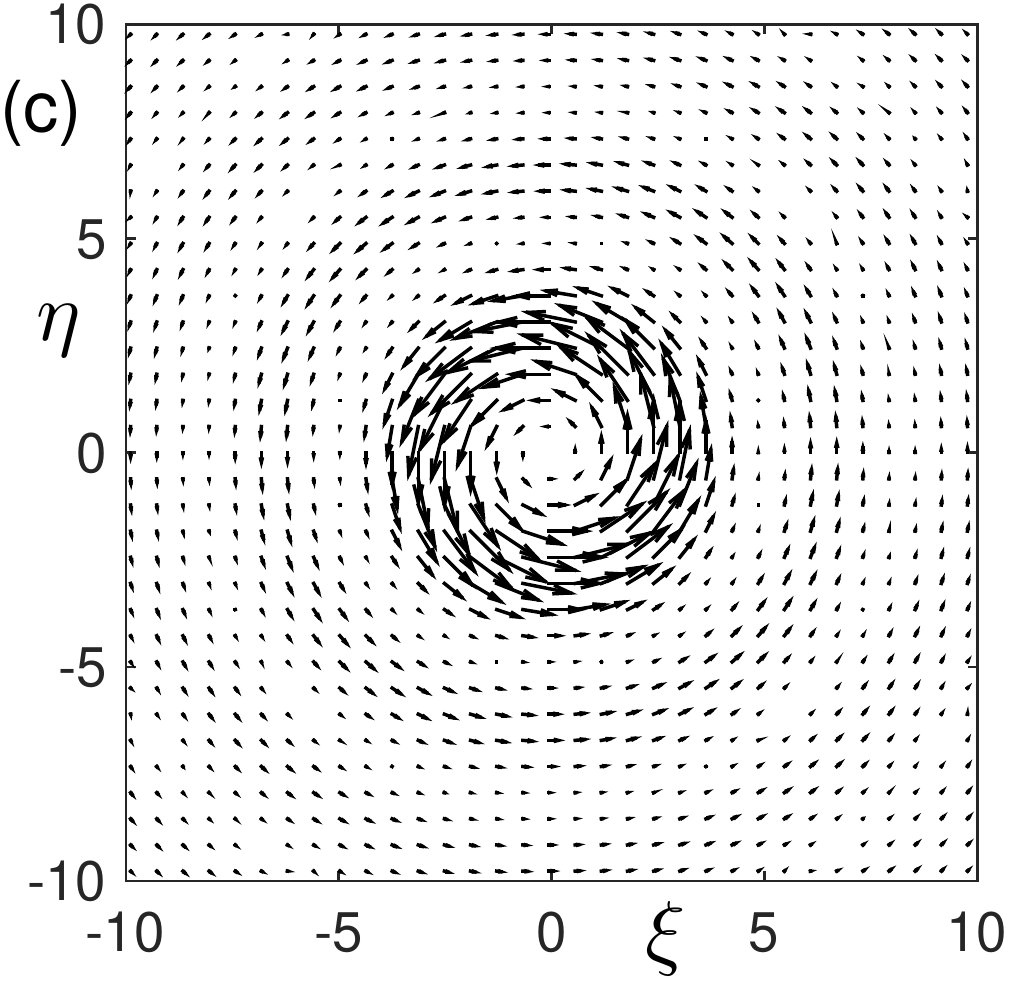}\includegraphics*[width=4.3cm]{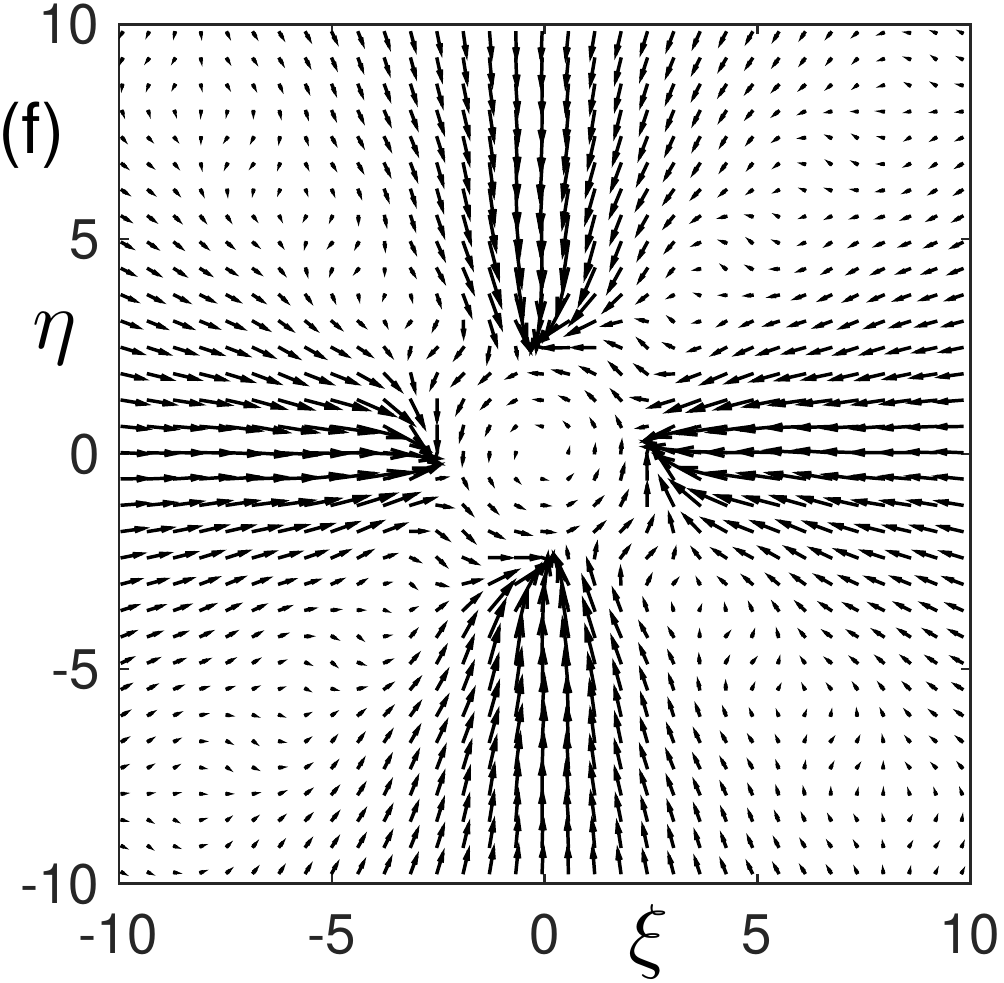}
\end{center}
\caption{\label{Fig7} (a) Intensity, (b) phase and (c) intensity current of the initial superposition of two Bessel beams of opposite charges, $s=2$ and $-s=-2$, average amplitude $b_L=3$ and azimuthal contrast $C_L=0.98$ ($b_{L,s}=2.323$ and $b_{L,-s}=1.899$). (d) Intensity, (e) phase and (f) intensity current of the final stationary state in the form of a dissipaton formed in a medium with $M=5$ and $\alpha\simeq 0$. The inset in (d) is the dissipation pattern $\tilde a^{2M}$.}
\end{figure}

Of particular interest are the dissipatons with $2s$-fold rotational symmetry formed from two Bessel beams
\begin{equation}
\tilde A(\rho,\varphi,0)=b_{L,s} J_s(\rho)e^{is\varphi}+ b_{L,-s}J_{-s}(\rho)e^{-is\varphi}
\end{equation}
of opposite topological charges $\pm s$. Without loss of generality, we can choose real and positive $b_{L,s}$ and $b_{L,-s}$, since the only effect of a phase shift between them is a rigid beam rotation. It is also convenient to fix their values as $b_{L,s}^2=b_L^2[1+\sqrt{1-C_L^2}]/2$ and $b_{L,-s}^2=b_L^2[1-\sqrt{1-C_L^2}]/2$, with $0\le C_L\le 1$, yielding the circularly symmetric case $b_LJ_s(\rho)e^{is\varphi}$ for $C_L=0$ ($b_{L,s}=b_L$, $b_{L,-s}=0$). The choice $C_L>0$ ($b_{L,s}>b_{L-s}$) introduces an azimuthal modulation of period $\pi/s$ and depth $C_L$ in the intensity pattern, but it continues to have a vortex of charge $s$ in the origin. In the limit case with $C_L=1$ ($b_{L,s}=b_{L,-s}$), the intensity vanishes at $2s$ azimuthal directions. Two examples of these input beams with respective low and high modulation depths ($C_L=0.3$ and $C_L=0.98$) are shown in Figs. \ref{Fig6}(a-c) and \ref{Fig7}(a-c) for $s=2$. With increasing $C_L$, the azimuthal change in $2\pi s$ of the phase about the vortex approaches an step-like variation, reaching $2s$ discontinuous jumps of $\pi$ in the limit $C_L=1$, the intensity current $\mathbf j = (s/\rho)(b_{L,s}^2-b_{L,-s}^2)J_s^2(\rho)\mathbf u_\varphi$ is always purely azimuthal, but increasingly weaker, and the associated angular momentum density $L=\tilde a^2 \partial \Phi/\partial\varphi=\mbox{Im}\{\tilde A^\star \partial \tilde A/ \partial \varphi\}=s (b_{L,s}^2-b_{L,-s}^2)J_s^2(\rho)$ diminishes, vanishing in the limit $C_L=1$. We focus on these linear Bessel superpositions as input beams because the formed dissipatons share a number of their properties, but we stress that the same dissipatons are formed from other beams with equivalent asymptotic inward currents, that is, from $\tilde A(\rho,\varphi,0) = b_{L,s} H_s^{(2)}(\rho)e^{is\varphi}+ b_{L,-s}H_{-s}^{(2)}(\rho)e^{-is\varphi} + f(\rho,\varphi)$, where $f(\rho,\varphi)$ does not carry any additional inward radial current at large radius, that do not have any of these properties around the beam center. There may not even be light in the central region of the input beam.

The intensity, phase and current patterns of the respective dissipatons formed after a transient propagation stage are depicted in Figs. \ref{Fig6} (d-f) and Figs. \ref{Fig7} (d-f). The corresponding dissipation patterns with four localized dissipation spots forming a square are shown in the insets of Figs. \ref{Fig6} (d) and \ref{Fig7} (d). The dissipathons always preserve the $2s$-fold rotational symmetry of the input Bessel beam superposition, the vortex of the same charge $s$ at $\rho=0$, and an azimuthal modulation of the intensity profile of similar depth as that of the input beam. We stress that symmetrization into a circularly symmetric nonlinear BVB is never observed, irrespective of the smallness of $C_L$ of the input beam. As seen below, this is a consequence of the conservation of the inward H\"ankel amplitudes. The step-like azimuthal variation of the phase profiles also preserves approximately the same sharpness as that of the input beam, although it is slightly changing with radial distance $\rho$. In the limit of an input beam with $C_L=1$, the formed dissipaton continues to have $2s$ null directions, and the azimuthal variation of the phase $2s$ discontinuous jumps. In all cases, a substantial difference from the input beam is that the azimuthal variation of the phase is accompanied by a pronounced radial variation, associated with the permanently converging wave fronts and the permanent radial inward component of the intensity current sustaining the stationarity. For increasing azimuthal input contrast, the radial inward component is accompanied by decreasing azimuthal component and angular momentum, that vanish in the limit $C_L=1$.

\begin{figure}
\begin{center}
\includegraphics*[width=4.3cm]{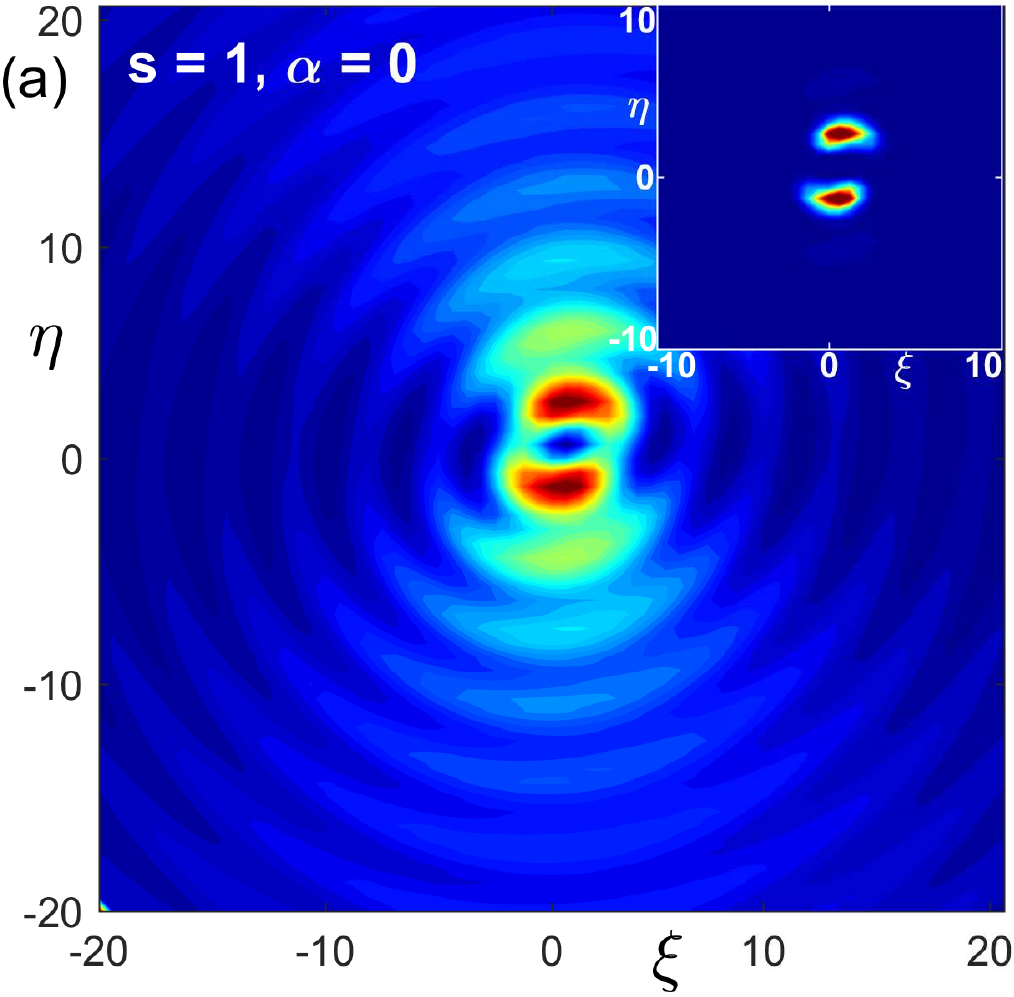}\includegraphics*[width=4.3cm]{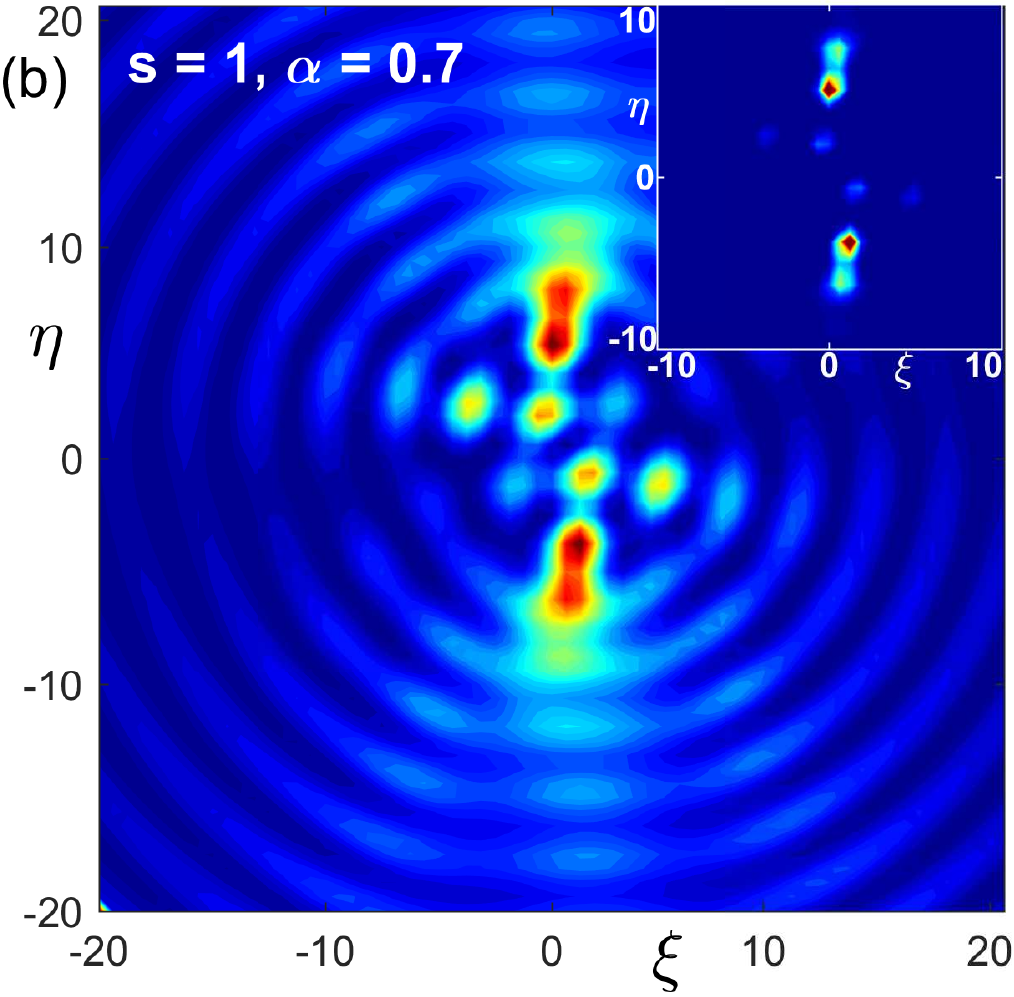}
\includegraphics*[width=4.3cm]{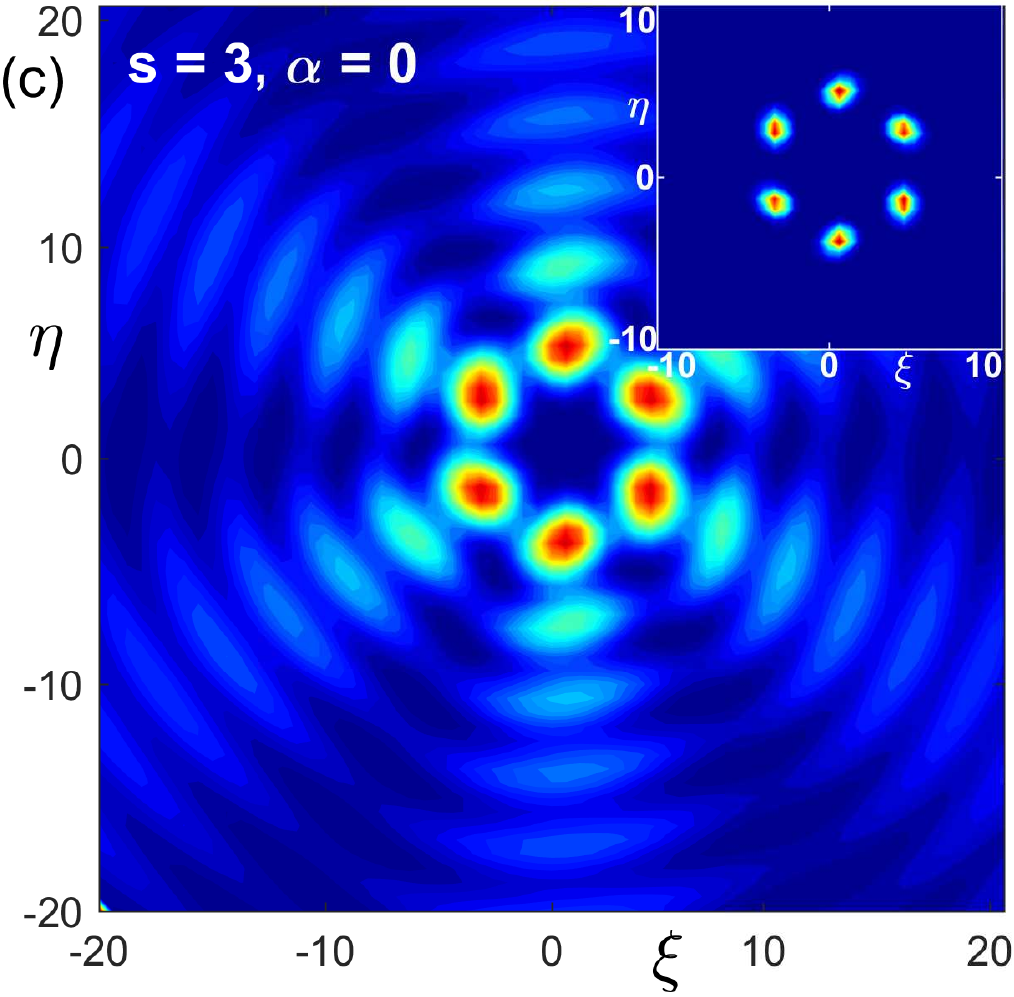}\includegraphics*[width=4.3cm]{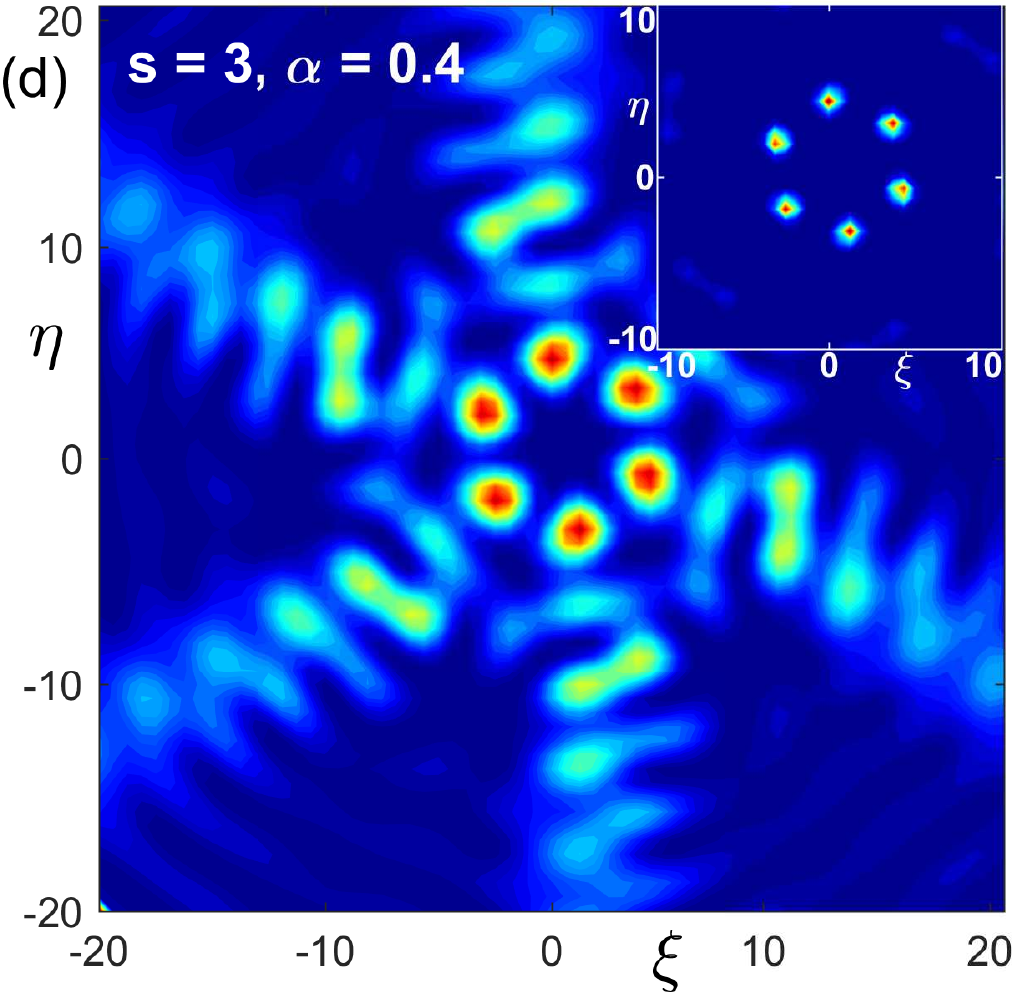}
\end{center}
\caption{\label{Fig8} (a) and (b) Intensity and dissipation patterns (insets) of the dissipahons formed starting from two conical beam superposition with average amplitude $b_L=3$, contrast $C=0.8$ ($b_{L,s}=2.683$ and $b_{L,-s}=1.342$), and $s=1$ in media with $M=5$ and cone angles and Kerr nonlinearity such that $\alpha\simeq 0$ and $\alpha=\alpha_{\rm max}$. (c-d) The same except that $s=3$.}
\end{figure}

In the above examples, we chose $\alpha\simeq 0$ for simplicity, but similar properties hold for $0\le\alpha\le\alpha_{\rm max}$, where $\alpha_{\rm max}$ depends now on $s$, $b_L$, $C_L$ and $M$. For $\pm s=\pm 1$ and $\pm s=\pm 3$, the intensity pattern of the dissipatons with $2$-fold and $6$-fold rotational symmetry formed in a medium with $M=5$ and cone angles and Kerr nonlinearities such that $\alpha\simeq 0$ and $\alpha_{\rm max}$ are depicted in Fig. \ref{Fig8}. The insets show the corresponding two-spot and six-spot dissipation patterns. We note that even if the intensity profile is weakly localized and becomes increasingly irregular with increasing $\alpha$, these dissipation patterns continue to be extremely localized and ordered.

\begin{figure}
\begin{center}
\includegraphics*[width=9cm]{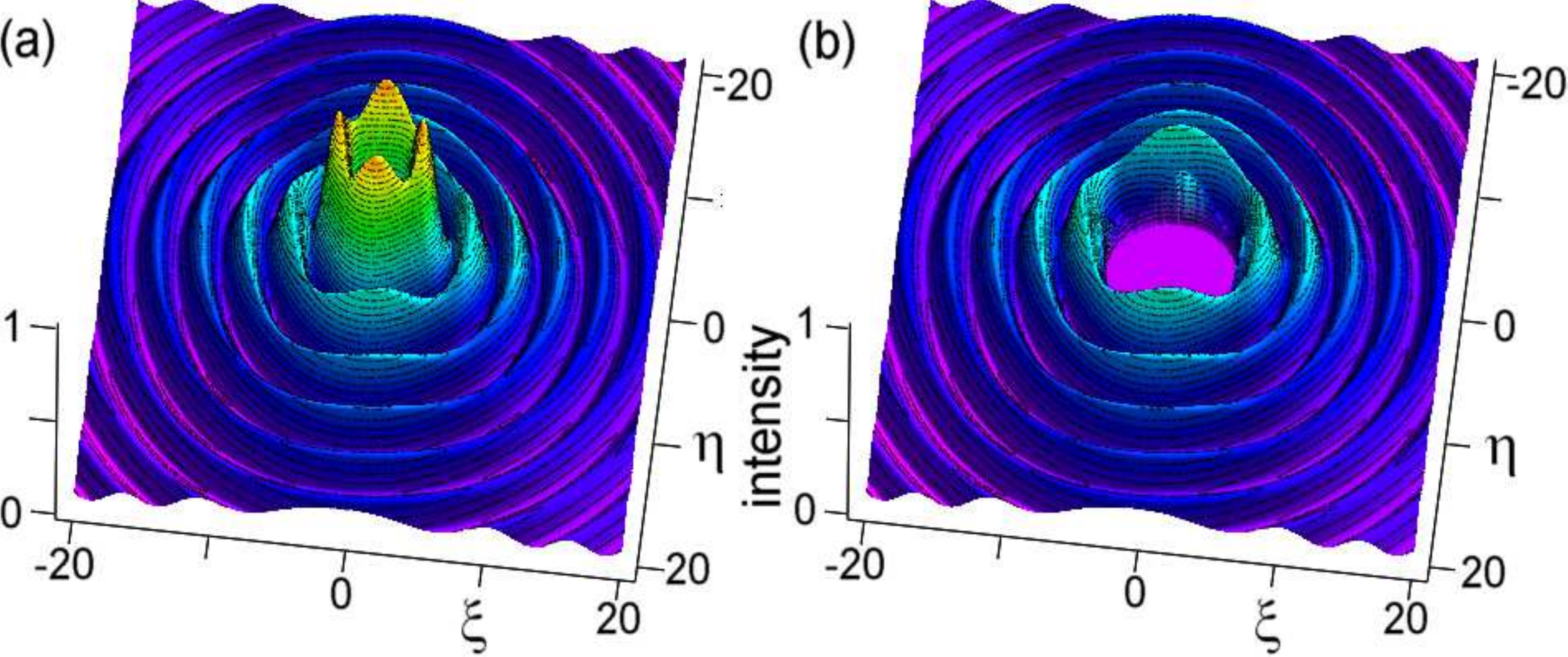}
\includegraphics*[width=4.5cm]{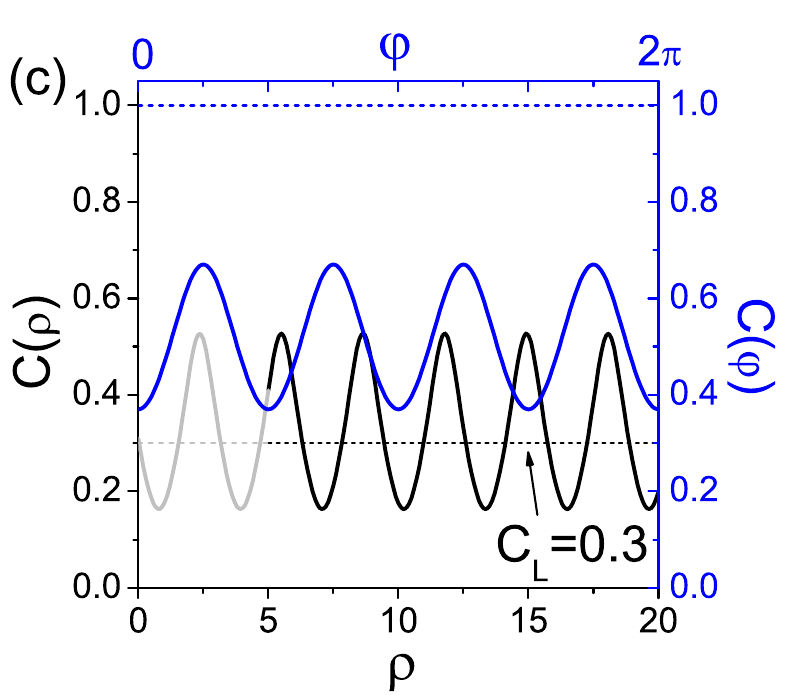}\includegraphics*[width=4cm]{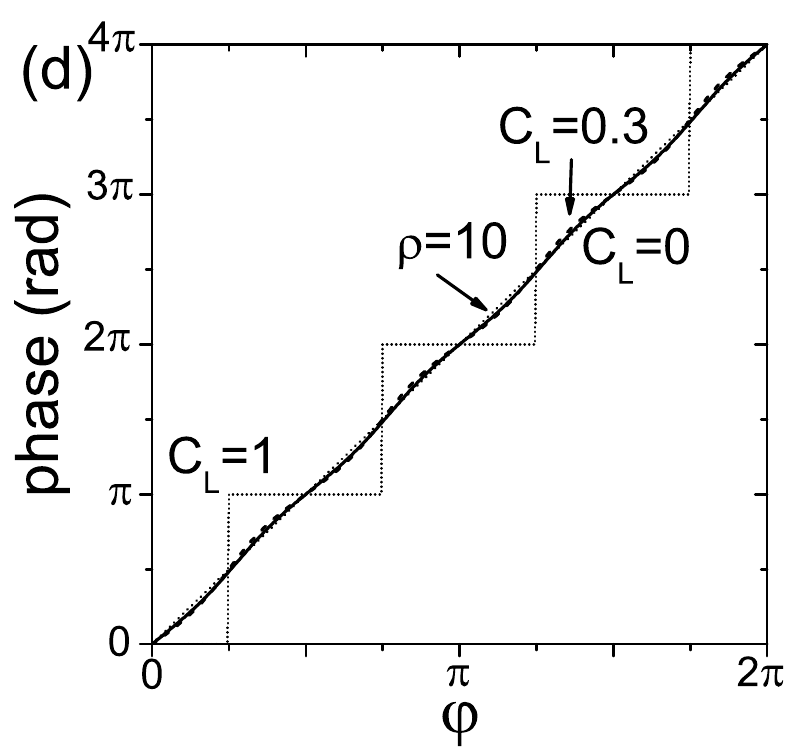}
\end{center}
\caption{\label{Fig9} (a) and (b) For the dissipaton in Fig. \ref{Fig6} formed from the input Bessel beam superposition with $C_L=0.3$, its intensity profile and its asymptotic form from Eq. (\ref{ASYMP3}) with $b_{\rm in,s}=b_{L,s}=2.965$, $b_{\rm in,-s}=b_{L,-s}=0.455$, $b_{\rm out,s}=0.811$, $b_{\rm out,-s}=-0.153$, verifying $|b_{{\rm in},s}|^2+|b_{{\rm in},-s}|^2-|b_{{\rm out},s}|^2-|b_{{\rm out},-s}|^2=N_\infty=8.34$. (c) Dashed and solid blue curves: Contrast of the radial oscillations at each azimuthal direction, evaluated from the above H\"ankel amplitudes and Eqs. (\ref{BPHI2}) Eq. (\ref{CPHI}) for the input Bessel beam superposition and for the dissipaton. Dashed and solid black curves: Contrast of the azimuthal oscillations at each radius, obtained from the H\"ankel amplitudes and Eqs. (\ref{BRHO}) and (\ref{CRHO}) for the input Bessel beam superposition (equal to $C_L=0.3$) and for the dissipaton. (d) Azimuthal variation of the phase at the indicated radius, obtained from the H\"ankel amplitudes, and Eqs (\ref{BRHO}) and (\ref{PHAS}). The dashed curve represents the same quantity for the input beam with $C_L=0.3$, and the dotted curves the azimuthal variation of the phase in the limit cases of input beams with $C_L=0$ and $C_L=1$, in which case the discontinuous jumps of the phase are preserved.}
\end{figure}

\begin{figure}
\begin{center}
\includegraphics*[width=9cm]{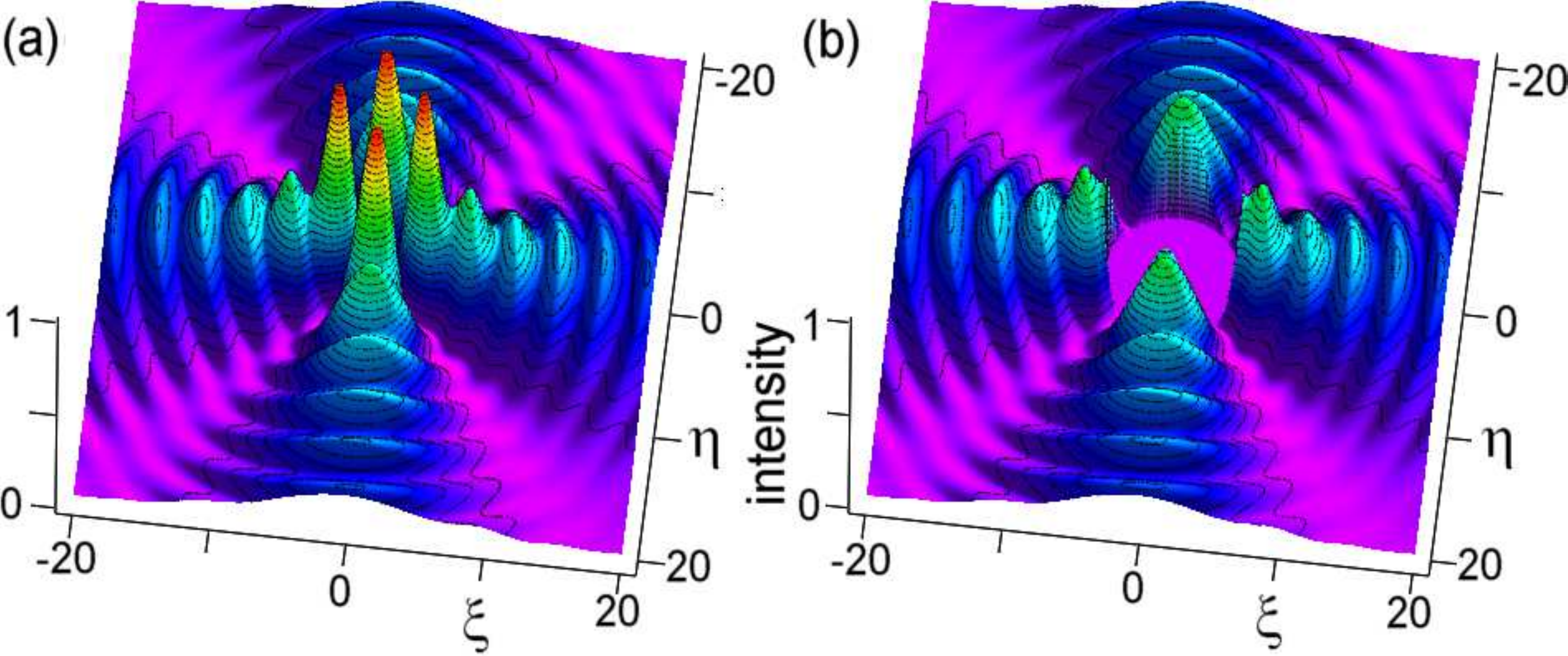}
\includegraphics*[width=4.5cm]{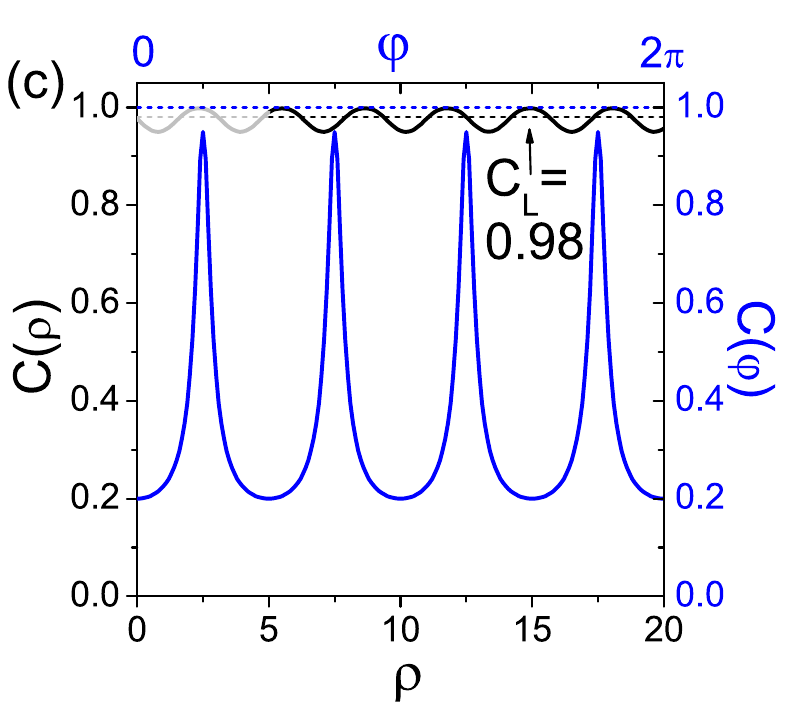}\includegraphics*[width=4cm]{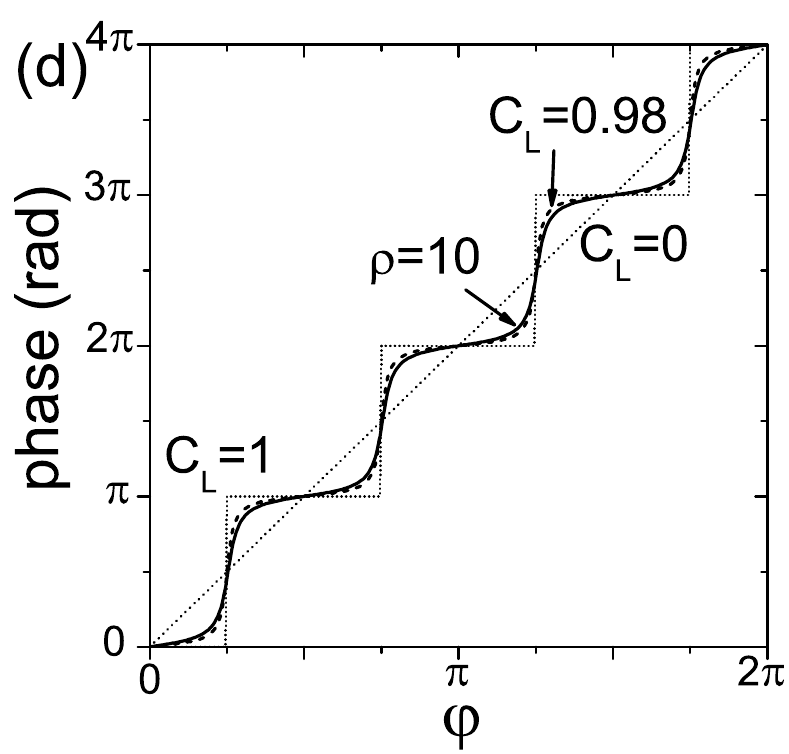}
\end{center}
\caption{\label{Fig10} (a) and (b) For the dissipaton in Fig. \ref{Fig7} formed from the input Bessel beam superposition with $C_L=0.98$, its intensity profile and its asymptotic form from Eq. (\ref{ASYMP3}) with $b_{\rm in,s}=b_{L,s}=2.323$, $b_{\rm in,-s}=b_{L,-s}=1.899$, $b_{\rm out,s}=0.367$, $b_{\rm out,-s}=0.057$, verifying $|b_{{\rm in},s}|^2+|b_{{\rm in},-s}|^2-|b_{{\rm out},s}|^2-|b_{{\rm out},-s}|^2=N_\infty=8.76$. (c) Dashed and solid blue curves: Contrast of the radial oscillations at each azimuthal direction, evaluated from the above H\"ankel amplitudes and Eqs. (\ref{BPHI2}) Eq. (\ref{CPHI}) for the input Bessel beam superposition and for the dissipaton. Dashed and solid black curves: Contrast of the azimuthal oscillations at each radius, obtained from the H\"ankel amplitudes and Eqs. (\ref{BRHO}) and (\ref{CRHO}) for the input Bessel beam superposition (equal to $C_L=0.98$) and for the dissipaton. (d) Azimuthal variation of the phase at the indicated radius, obtained from the H\"ankel amplitudes, and Eqs (\ref{BRHO}) and (\ref{PHAS}). The dashed curve represents the same quantity for the input beam with $C_L=0.98$. The dotted curves are as in Fig. \ref{Fig9}.}
\end{figure}

The above structure of these dissipatons is reflected in the structure beyond the dissipative region, where it admits an analytical description. The asymptotic form in Eq. (\ref{ASYMP1}) for these dissipatons reads as
\begin{eqnarray}\label{ASYMP3}
\tilde A &\simeq& \frac{1}{2}\left\{\left[b_{{\rm out},s}H_{s}^{(1)}(\rho)+ b_{{\rm in},s}H_{s}^{(2)}(\rho)\right]e^{is\varphi}\right. \nonumber \\
&+& \left.\left[b_{{\rm out},-s}H_{-s}^{(1)}(\rho)\!+\! b_{{\rm in},-s}H_{-s}^{(2)}(\rho)\right]e^{-is\varphi}\!\right\} e^{-i\zeta}\, .
\end{eqnarray}
Equivalently, Eq. (\ref{ASYMP2}) also holds, where now
\begin{eqnarray}\label{BPHI2}
b_{\rm out}(\varphi) &=& b_{{\rm out},s} e^{is\left(\varphi-\frac{\pi}{2}\right)}+ b_{{\rm out},-s}e^{-is\left(\varphi-\frac{\pi}{2}\right)}\,, \\
b_{\rm in} (\varphi) &=& b_{{\rm in},s} e^{is\left(\varphi+\frac{\pi}{2}\right)}+ b_{{\rm in},-s}e^{-is\left(\varphi+\frac{\pi}{2}\right)}\, ,
\end{eqnarray}
represent harmonic azimuthal oscillations of period $2\pi/s$. Consequently, the contrast $C(\varphi)$ of the radial oscillations in the intensity profile along each azimuthal angle $\varphi$, and its average value $R(\varphi)$, repeat themselves each period $\pi/s$, as expected. For the dissipatons in Figs. \ref{Fig6} and \ref{Fig7}, Figs. \ref{Fig9} (a,b) and \ref{Fig10} (a,b) depict the intensity profiles and their asymptotic forms satisfying the conservation of the inward H\"ankel amplitudes, $|b_{\rm in,s}|=b_{L,s}$, $|b_{\rm in,-s}|=b_{L,-s}$, and the replenishment condition, $|b_{\rm in,s}|^2+|b_{\rm in,-s}|^2- |b_{\rm out,s}|-|b_{\rm out,-s}|^2=N_\infty$. From these values, the contrast of the radial oscillations at each azimuthal direction, $C(\varphi)$, are evaluated from Eqs. (\ref{BPHI2}) and (\ref{CPHI}), and are plotted in Figs. \ref{Fig9} (c) and \ref{Fig10}(c) as solid blue curves. As for nonlinear BVBs, dissipation manifests in a loss of radial oscillation contrast (from unity for the input beam), with the peculiarity that the decrease depends on the azimuthal direction and is more pronounced along those of higher intensity (e. g., at $\varphi=0$).

For dissipatons with $2s$-fold rotational symmetry, using the asymptotic forms of H\"ankel functions in Eq. (\ref{HANKEL}) again, but grouping now the terms with the same azimuthal dependence, Eq. (\ref{ASYMP3}) we can alternatively be written as
\begin{equation}\label{ASYMP4}
\tilde A= \frac{1}{\sqrt{2\pi\rho}}\left[b_s(\rho)e^{is\varphi}+ b_{-s}(\rho)e^{-is\varphi}\right]e^{-i\zeta} \, ,
\end{equation}
where we have defined
\begin{eqnarray}\label{BRHO}
b_s(\rho) &=& b_{{\rm out},s}e^{i\left(\rho-\frac{\pi}{4}-\frac{\pi s}{2}\right)}+b_{{\rm in},s}e^{-i\left(\rho-\frac{\pi}{4}-\frac{\pi s}{2}\right)}   \, , \\
b_{-s}(\rho) &=& b_{{\rm out},-s}e^{i\left(\rho-\frac{\pi}{4}+\frac{\pi s}{2}\right)}\!+\!b_{{\rm in},-s}e^{-i\left(\rho-\frac{\pi}{4}+\frac{\pi s}{2}\right)}  \, .
\end{eqnarray}
In this form, the intensity profile at large radius is also given by
\begin{eqnarray}\label{AZIMUTALI}
2\pi \rho |\tilde A|^2\!&\simeq&\! |b_s(\rho)|^2\!+\!|b_{-s}(\rho)|^2\!+\!2|b_s(\rho)||b_{-s}(\rho)|  \nonumber\\
     &\times &
\cos\left[2s \varphi \! +\!\mbox{arg}\,b_s(\rho)\!-\!\mbox{arg}\,b_{-s}(\rho)\right]\,.
\end{eqnarray}
As a function of $\varphi$, these are harmonic oscillations of period $\pi/s$ of $\rho$-dependent contrast
\begin{equation}\label{CRHO}
C(\rho)=\frac{2|b_s(\rho)||b_{-s}(\rho)|}{|b_s(\rho)|^2+|b_{-s}(\rho)|^2}
\end{equation}
about a $\rho$-dependent average value $R(\rho) =|b_s(\rho)|^2+|b_{-s}(\rho)|^2$. The azimuthal contrast as a function of the radial distance for the dissipatons in Figs. \ref{Fig9} (a) and \ref{Fig10}(a) are plotted as black curves in Figs. \ref{Fig9}(c) and \ref{Fig10}(c). As seen, the azimuthal contrast, which was independent of $\rho$ and equal to $C_L$ for the input Bessel beam superposition, oscillates with radial distance, remaining however of the same order as for the input beam, $C_L$, as observed from the numerical simulations. In the particular dissipatons formed from  $C_L=1$ ($b_{L,s}=b_{-L,s}=b_L/\sqrt{2}$) symmetry considerations ($b_{{\rm in},s}=b_{{\rm in},-s}$, and $b_{{\rm, out},s}=b_{{\rm out},-s}$) imply $b_{-s}(\rho)=(-1)^sb_s(\rho)$, and therefore $C(\rho)=1$ too, i. e., the dissipaton preserves the $2s$ directions of zero intensity of the input Bessel beam superposition.

In addition to the inward radial component of the intensity flux, these dissipatons with $2s$-fold symmetry carry the azimuthal component  $j_\varphi\simeq (s/2\pi\rho)[|b_s(\rho)|^2-|b_{-s}(\rho)|^2]/\rho$ at large radius. Interestingly, while the intensity and phase profiles exhibit  pronounced azimuthal variations, the azimuthal flux is circularly symmetric. Thus, the vector expression of the intensity current is
\begin{equation}
\mathbf j\simeq \frac{[|b_{\rm out}(\varphi)|^2\!-\!|b_{\rm in}(\varphi)|^2]}{2\pi\rho}\mathbf u_\rho
                                  +  \frac{s}{\rho}\frac{[|b_s(\rho)|^2\! -\!|b_{-s}(\rho)|^2]}{2\pi\rho}\mathbf u_\varphi.
\end{equation}
Accordingly, the angular momentum density is given by
\begin{equation}
L= \mbox{Im}\left\{\tilde A^\star \frac{\partial \tilde A}{\partial \varphi}\right\}\simeq \frac{s}{2\pi\rho} [|b_s(\rho)|^2 - |b_{-s}(\rho)|^2]\, ,
\end{equation}
and is also circularly symmetric. These properties are also of interest in order to evaluate the azimuthal derivative of the phase as $\partial\Phi/\partial\varphi = L/|\tilde A|^2$. Upon integration with the above asymptotic expressions of $L$ and $|\tilde A|^2$, the phase is asymptotically given by
\begin{equation}\label{PHAS}
\Phi\simeq \tan^{-1}\left\{\frac{|b_s(\rho)|-|b_{-s}(\rho)|}{|b_s(\rho)|+|b_{-s}(\rho)|}\tan\left[s\varphi+ \kappa(\rho)\right]\right\}\, ,
\end{equation}
with $\kappa(\rho)=[\mbox{arg}\,b_s(\rho)-\mbox{arg}\, b_{-s}(\rho)]/2$, and aside the term $-\zeta$ on propagation. From Eqs. (\ref{BRHO}) and (\ref{CRHO}), the above relation can also be expressed as
\begin{equation}
\Phi\simeq \tan^{-1}\left\{\sqrt{\frac{1-C(\rho)}{1+C(\rho)}}\tan[s\varphi + \kappa(\rho)]\right\}\, ,
\end{equation}
relating the azimuthal variation of the phase with that of the intensity. The azimuthal phase variation of the dissipatons of Figs. \ref{Fig9} (a) and \ref{Fig10} (a) are plotted in the respective Figs. \ref{Fig9} (d) and \ref{Fig10} (d) at a particular radius. These explain the increasing sharpness of the azimuthal phase variation with increasing azimuthal contrast of the intensity $C(\rho)$, and that in the limit $C(\rho)=1$ (obtained from $C_L=1$), the phase of the azimuthons continues to display $2s$ discontinuous $\pi$ jumps of the input Bessel beam superposition.

Dissipatons with $2s$-fold rotational symmetry share with azimuthons, and their limiting cases of vortex solitons and soliton clusters, similar azimuthal modulation of the intensity patterns and similar stair-like azimuthal variation of the phase \cite{DESYATNIKOVPrl1,MINOVICHOe}, but as pointed out in the introduction, they are supported by completely different mechanisms. Also, azimuthons are generally rotating, and the number of azimuthal maxima and vortex charge in their center are independent properties, while these dissipathons are static, and the number of maxima is $2s$ if the charge is $s$.

\section{Conclusions}

In conclusion, we have described a broad family of propagation-invariant and robust conical beams in media with Kerr-type nonlinearities and nonlinear absorption, with the property of continuously dissipating power along narrow and ordered channels, whose number and geometrical disposition can be easily controlled. We have described semi-analytically their spatial structure and found the laws that govern their formation from superpositions of Bessel beams.

In filamentation experiments, the new degrees of control offered by these beams with regard to the number and disposition of plasma channels may open new possibilities in short-range applications such as material laser ablation, wave guide writing, and in long-range applications such as filamentation guiding of weak waves in the atmosphere \cite{COURVOISIERApplPhys,BHUYAN,KRISNALaserPhysics,ARNOLDJPhysB,COURVOISIER2}.

M.A.P. acknowledges support from Projects of the Spanish Ministerio de Econom\'{\i}a y Competitividad No. MTM2012-39101-C02-01, MTM2015-63914-P, and No. FIS2013-41709-P.


\begin{thebibliography}{99}

\bibitem{MALOMED1} B. A. Malomed, D. Mihalache, F. Wise, and L. Torner, Spatiotemporal optical solitons, J. Opt. B {\bf7,} R53 (2000).
\bibitem{KIVSHAR} Y. S. Kivshar and G. P. Agrawal, {\em Optical Solitons: From Fibers to Photonic Crystals} (Academic Press, San Diego, 2003).
\bibitem{CHEN} Z. Chen, M. Segev, and D. N Christodoulides, Optical spatial solitons: historical overview and recent advances, Rep. Prog. Phys. {\bf 75,} 086401 (2012).
\bibitem{AKHMEDIEV} N. Akhmediev and A. Ankiewicz (Eds.) {\em Dissipative solitons}, Lecture Notes in Physics {\bf 661,}  (Springer-Verlag, Berlin Heidelberg, 2005).
\bibitem{ACKEMANN} T. Ackemann, W. J. Firth, G-L Oppo, Fundamentals and applications of spatial dissipative solitons in photonic devices, Advances In Atomic, Molecular, and Optical Physics {\bf 57,} 323–421 (2009).
\bibitem{MALOMED2} B. A. Malomed, Spatial solitons supported by localized gain, J. Opt. Soc. Am. B {\bf 31,} 2460-2475 (2014).
\bibitem{KRUGLOV}  V. I. Kruglov and R. A. Vlasov, Spiral self-trapping propagation of optical beams in media with cubic nonlinearity, Phys. Lett. A {\bf 111,} 401 (1985).
\bibitem{DESYATNIKOVPrl1} A. S. Desyatnikov, A. A. Sukhorukov, and Yu. S. Kivshar, Azimuthons: spatially modulated vortex solitons, Phys. Rev. Lett. {\bf 95,} 203904 (2005).
\bibitem{MINOVICHOe} A. Minovich, D. N. Neshev, A. S. Desyatnikov, W. Krolikowski, and Y. S. Kivshar, Observation of optical azimuthons, Opt. Express {\bf 17,} 23610 (2009).
\bibitem{DESYATNIKOVPrl2} A. S. Desyatnikov and Yu. S. Kivshar, Rotating optical soliton clusters, Phys. Rev. Lett. {\bf 88,} 053901 (2002).
\bibitem{DESYATNIKOVReview} A. S. Desyatnikov, L. Torner, and Y. S. Kivshar, Optical vortices and vortex solitons, Prog. Opt. {\bf 47,} 291 (2005).
\bibitem{PAULAU} P. V. Paulau, D. Gomila, P. Colet, N. A. Loiko, N. N. Rosanov, T. Ackemann, and W. J. Firth, Vortex solitons in lasers with feedback, J. Opt. Express {\bf 18,} 8859-8866 (2010).
\bibitem{CRASOVAN} L.C. Crasovan, B. A. Malomed, and D. Mihalache, Stable vortex solitons in the two-dimensional Ginzburg-Landau equation, Phys. Rev. E {\bf 63,} 016605 (2000).
\bibitem{ZHANGPra} Y. Zhang, M. Belic, Z. Wu, C. Yuan, R. Wang, K. Lu, and Y. Zhang, Multicharged optical vortices induced in a dissipative atomic vapor system, Phys. Rev. A {\bf 88,} 013847 (2013).
\bibitem{PORRASPrl} M. A. Porras, A. Parola, D. Faccio, A. Dubietis, and P. Di Trapani, Nonlinear Unbalanced Bessel Beams: Stationary
ConicalWaves Supported by Nonlinear Losses, Phys. Rev. Lett. {\bf 93,} 153902 (2004).
\bibitem{COUAIRONPhysRep} A. Couairon and A. Mysyrowicz, Femtosecond filamentation in transparent media, Phys. Rep. {\bf 441,} 47--189 (2007).
\bibitem{BERGERep} L. Berg\'e, S. Skupin, R. Nuter, J. Kasparian, and J. P. Wolf, ``Ultrahsort filaments of light in weakly ionized, optically transparent media," Reports on Progress in Physics {\bf 70,} 109801 (2008).
\bibitem{BERGEDis} L. Berg\'e and S Skupin, Modeling ultrashort filaments of light, Discrete and Continuous Dynamical Systems {\bf 23,} 1099-1139 (2009).
\bibitem{DUBIETIS} A. Dubietis, E. Kucinskas, G. Tamosauskas, E. Gaizauskas, M. A. Porras, and P. Di Trapani, Self-reconstruction of light filaments, Opt. Lett. {\bf 24,} 2893-2895 (2004).
\bibitem{GAIZAUSKAS} E. Gaizauskas, A. Dubietis, V. Kudriasov, V. Sirutkaitis, A. Couairon, D. Faccio, P. Di Trapani, On the role of Conical Waves in Self-focusing and filamentation of femtosecond pulses with nonlinear losses, Top. Appl. Phys. {\bf 114,} 457-479 (2009).
\bibitem{FACCIOOptComm} D. Faccio, M. A. Porras, A. Dubietis, G. Tamosauskas, E. Kucinskas, A. Couairon, and P. Di Trapani, Angular and chromatic dispersion in Kerr-driven conical emission, Opt. Commun. {\bf 265,} 672-677 (2006).
\bibitem{ROSOAppPhysA} L. Roso, J. San Rom\'an, I. J. Sola, C. Ruiz, V. Collados, J. A. P\'erez, C. M\'endez, J. R. V\'azquez de Aldana, I. Arias, L. Plaja, Propagation of terawatt laser pulses in the air, Appl. Phys. A {\bf 92,} 865 (2008).
\bibitem{ZEMLYANOV} A. A. Zemlyanov and A. D. Bulygin, Features of the development of light field perturbations in a kerr medium with nonliner absorption, Atmospheric and Oceanic Optics {\bf 26,} 85-89 (2013).
\bibitem{DUBIETISLithuanian} A. Dubietis, G. Valiulis, and A. Varanavicius, Nonlinear localization of light, Lithuanian Journal of Physics {\bf 146,} 7-18 (2006).
\bibitem{FACCIOPra} D. Faccio, A. Matijosius, A. Dubietis, R. Piskarskas, A. Varanavičius, E. Gaizauskas, A. Piskarskas, A. Couairon,
and P. Di Trapani, Near- and far-field evolution of laser pulse filaments in Kerr media, Phys. Rev. E {\bf 72,}  037601 (2005).
\bibitem{DURNINPrl} J. Durnin, J. J. Miceli, and J. H. Eberly, Diffraction-free beams, Phys. Rev. Lett {\bf 58,} 1499-1501 (1987).
\bibitem{SIVILOGLOUOptLett} G. A. Siviloglou, J. Broky, A. Dogariu, D. N. Christodoulides, Observation of accelerating Airy beams, Phys. Rev. Lett {\bf 99,} (2007).
\bibitem{PORRASJosaB} M. A. Porras and C. Ruiz-Jim\'enez, Nondiffracting and nonattenuating vortex light beams in media with nonlinear absorption of orbital angular momentum, J. Opt. Soc. Am B {\bf 31,} 2657-2664 (2014).
\bibitem{JUKNAOe} V. Jukna, Mili\'an, C. Xie, T. Itina, J. Dudley, F. Courvoisier, and A. Couairon, Filamentation with nonlinear Bessel vortices,
Opt. Express {\bf 22,} 25410 (2014).
\bibitem{XIESci} C. Xie, V. Jukna, C. Mili\'an, R. Giust, I. Ouadghiri-Idrissi, T. Itina, J. M. Dudley, A.Couairon, and F.Courvoisier, Tubular filamentation for laser material processing, Sci. Rep. {\bf 5,} 8914 (2015).
\bibitem{LOTTIPra} A. Lotti, D. Faccio, A. Couairon, D. G. Papazoglou, P. Panagiotopoulos, D. Abdollahpour, and S. Tzortzakis, Stationary
nonlinear Airy beams, Phys. Rev. A {\bf 84,} 021807(R) (2011).
\bibitem{POLESANAPra} P. Polesana, M. Franco, A. Couairon, D. Faccio, and P. Di Trapani, Phys. Rev. A {\bf 77,} 043814 (2008).
\bibitem{AKTURKOptComm} S. Akturk, B. Zhou, M. Franco, A. Couairon, A. Mysyrowicz, Generation of long plasma channels in air by focusing ultrashort laser pulses with an axicon, Opt. Commun {\bf 282,} 129–134 (2009).
\bibitem{POLESANAPre} P. Polesana, A. Dubietis, M. A. Porras, E. Kucinskas, D. Faccio, A. Couairon, and P. Di Trapani, Near-field dynamics of ultrashort pulsed Bessel beams in media with Kerr nonlinearity, Phys. Rev. E {\bf 73,} 056612 (2006).
\bibitem{POLESANAPrl} P. Polesana, A. Couairon, D. Faccio, A. Parola, M. A. Porras, A. Dubietis, A. Piskarskas, and P. Di Trapani, Observation of Conical Waves in Focusing, Dispersive, and Dissipative Kerr Media, Phys. Rev. Lett {\bf 99,} 223902 (2007).
\bibitem{POLESANAOe} P. Polesana, D. Faccio, P. Di Trapani, A. Dubietis, A. Piskarskas, A. Couairon, and M. A. Porras, High localization, focal depth and contrast by means of nonlinear Bessel Beams, Opt. Express {\bf 13,} 6160-6167 (2005).
\bibitem{MAJUSJosaB} D. Majus and A. Dubietis, Statistical properties of ultrafast supercontinuum generated by femtosecond Gaussian and Bessel beams: a comparative study, J. Opt. Soc. Am. B {\bf 30,}  994--999 (2013).
\bibitem{AKTURKSpie} S. Akturk, B. Zhou, A. Houard, M. Franco, A. Couairon, and A. Mysyrowicz, Long plasma channels formed by axicon-focused filaments, Proc. SPIE 70271E (2008).
\bibitem{KAYAAip} G. Kaya, N. Kaya, M. Sayrac, Y. Boran, J. Strohaber, A. A. Kolomenskii,  M. Amani, and H. A. Schuessler, Extension of filament propagation in water with Bessel-Gaussian beams, AIP Advances {\bf 6,} 035001 (2016).
\bibitem{PORRASPra1} M. A. Porras, C. Ruiz-Jim\'enez, and J. C. Losada, Underlying conservation and stability laws in nonlinear propagation of axicon-generated Bessel beams, Phys. Rev. A {\bf 92}, 063826 (2015).
\bibitem{PORRASPra2} M. A. Porras, M\'arcio Carvalho, Herv\'e Leblond, and Boris A. Malomed, Stabilization of vortex beams in Kerr media by nonlinear absorption, Phys. Rev. A {\bf 94}, 053810 (2016).
\bibitem{COURVOISIERApplPhys} F. Courvoisier, J. Zhang, K. Bhuyan, M. Jacquot and J. M. Dudley, Applications of femtosecond Bessel beams to laser ablation, Appl. Phys. A {\bf 112,} 29-34 (2013).
\bibitem{BHUYAN} M. K. Bhuyan, P. K. Velpula, J. P. Colombier, T. Olivier, N. Faure, and R. Stoian, Single-shot high aspect ratio bulk nanostructuring of fused silica using chirp-controlled ultrafast laser Bessel beams, Appl. Phys. Lett. {\bf 104,} 021107 (2016).
\bibitem{KRISNALaserPhysics} G Krishna Podagatlapalli, S. Hamad, Md Ahamad Mohiddon, and S Venugopal Rao, Fabrication of nanoparticles and nanostructures using ultrafast laser ablation of silver with Bessel beams, Laser Physics Lett. {\bf 12,} 036003 (2015).
\bibitem{ARNOLDJPhysB} C. L. Arnold, S. Akturk, A. Mysyrowicz, V. Jukna, A. Couairon, T. Itina, R. Stoian, C. Xie, J. M. Dudley, F. Courvoisier,
Nonlinear Bessel vortex beams for applications,  J. Phys. B: At. Mol. Opt. Phys. {\bf 48,} 094006 (2015).
\bibitem{COURVOISIER2} F. Courvoisier, R. Stoian, A. Couairon, Ultrafast laser micro- and nano-processing with nondiffracting and curved beams, Optics \& Laser Technology {\bf 80,} 125–137 (2016).
\bibitem{FROEHLYJosaA} L. Froehly, M. Jacquot, P. A. Lacourt, J. M. Dudley, and F. Courvoisier, Spatiotemporal structure of femtosecond Bessel beams from spatial light modulators, J. Opt. Soc. Am. A {\bf 31,} 790-793 (2014).
\bibitem{CLARKOe} T. W. Clark, R. F. Offer, S. Franke-Arnold, A. S. Arnold, and N. Radwell, Comparison of beam generation techniques using a phase only spatial light modulator, Opt. Express {\bf 24,} 6249-6264 (2016).
\bibitem{VALILYEUOe} R. Vasilyeu, A. Dudley, N. Khilo, and A. Forbes, Generating superpositions of higher–order Bessel beams, Opt. Express {\bf 17,}  23389-23395 (2009).
\bibitem{ARLTOptCommun} J. Arlt and K. Dholakia, Generation of high-order Bessel beams by use of an axicon, Opt. Commun. {\bf 177,} 297–301 (2000).
\bibitem{COUAIRONSpie} A. Couairon, A. Lotti, P. Panagiotopoulos, D. Abdollahpour, D. Faccio, D. G. Papazoglou, S. Tzortzakis, F. Courvoisier, and
J. M. Dudley, Ultrashort pulse filamentation with Airy and Bessel beams, Proc. SPIE {\bf 8770,} 87701E (2013).
\end{thebibliography}
\end{document}